\documentclass[11pt]{article}
\usepackage[sc]{mathpazo} 
\usepackage{fullpage}
\usepackage[authoryear,sectionbib,sort]{natbib}
\usepackage[utf8]{inputenc}
\usepackage{lineno}
\usepackage{titlesec}
\usepackage{rotating}

\titleformat{\section}[block]{\Large\bfseries\filcenter}{\thesection}{1em}{}
\titleformat{\subsection}[block]{\Large\itshape\filcenter}{\thesubsection}{1em}{}
\titleformat{\subsubsection}[block]{\large\itshape}{\thesubsubsection}{1em}{}
\titleformat{\paragraph}[runin]{\itshape}{\theparagraph}{1em}{}[. ]

\usepackage{graphicx,amsmath,amssymb}

\title{The Minimum Environmental Perturbation Principle:\\ A New Perspective on Niche Theory}

\author{Robert Marsland III$^{1,\ast}$ \\ 
	Wenping Cui$^{1,2}$ \\ 
	Pankaj Mehta$^{1}$}

\date{}

\begin{document}
	
	\maketitle
	
	\noindent{} 1. Boston University, Boston, Massachusetts 02215;
	
	\noindent{} 2. Boston College, Chestnut Hill, Massachusetts 02467;
	
	\noindent{} $\ast$ Corresponding author; e-mail: marsland@bu.edu.

	
	\newpage{}
	
	\section*{Abstract}
	
	Fifty years ago, Robert MacArthur showed that stable equilibria optimize quadratic functions of the population sizes in several important ecological models. Here, we generalize this finding to a broader class of systems within the framework of contemporary niche theory, and precisely state the conditions under which an optimization principle (not necessarily quadratic) can be obtained. We show that conducting the optimization in the space of environmental states instead of population sizes leads to a universal and transparent physical interpretation of the objective function.  Specifically, the equilibrium state minimizes the perturbation of the environment induced by the presence of the competing species, subject to the constraint that no species has a positive net growth rate. We use this ``minimum environmental perturbation principle'' to make new predictions for eco-evolution and community assembly, and describe a simple experimental setting where its conditions of validity have been empirically tested.
	
	\newpage{}

\section*{Introduction}
The past century of research in theoretical ecology has revealed how simple mathematical models can have surprisingly rich behavior, with results that are often difficult to predict without running a numerical simulation. This is particularly the case when the number of simultaneously interacting species becomes large, and an exhaustive exploration of the parameter space is no longer possible. But deriving ecological insight from these models requires abstracting from an individual simulation run, to find qualitative features of the dynamics that generically follow from the basic modeling assumptions.

Fifty years ago, Robert MacArthur found hints of a general principle of this kind, concerning the properties of stable equilibrium states (\citealt{MacArthur1969,macarthur1970species}). In a model of competition for substitutable resources, now known by his name, he showed that the equilibrium states optimize a certain quadratic function of the population sizes. Under some additional assumptions, this function had a natural interpretation in terms of the difference between available resource production and the harvesting abilities of the consumers. He obtained similar optimization principles for several other models including one with direct interaction between resources and another representing competition to avoid predators, suggesting that this result might extend significantly beyond the specific context in which it was originally found. But he was unable to find an ecological interpretation of the objective function in these other cases, and no broader framework had yet been developed for systematically generalizing the principle.

In this paper, we complete MacArthur's work by situating it in the context of contemporary niche theory (cf. \citealt{chase2003ecological} for a thorough introduction). This mathematical and conceptual framework effectively generalizes the original consumer resource model to allow for arbitrary environmentally-mediated interactions, including saturating growth kinetics, competition for essential resources (e.g., as described by Liebig's Law of the Minimum), and microbial systems with rampant byproduct secretion. This framework first of all allows us to state the general conditions under which an optimization principle exists. But it also provides another benefit, by focusing our attention on the environmental state. Contemporary niche theory naturally lends itself to a graphical analysis in the space of environmental factors, where coexistence conditions can be geometrically determined (cf. \citealt{koffel2016geometrical} for a recent review). It turns out that conducting the optimization in this environmental space -- instead of in the space of population sizes -- leads to a generalizable ecological interpretation of the objective function. 

In the following sections, we first review MacArthur's original result, and describe how his model is generalized by the niche theory framework. Then we describe the general conditions for the existence of a optimization principle in a niche model, and show how the principle can be interpreted as a constrained minimization of the environmental perturbation induced by the competing species. We illustrate the scope of the result with seven examples: the three considered by MacArthur and four scenarios that depart from his assumptions in significant ways. One of these examples is taken from a classic experimental paper on resource competition in rotifer populations (\citealt{rothhaupt1988mechanistic}), where the model was shown to provide an excellent description of the experiments. We review how the conditions for an optimization principle can be directly verified in this case. Finally, we discuss an important corollary of our result, that the environmental perturbation monotonically increases during community assembly or evolution. 

\begin{figure*}
	\includegraphics[width=16cm]{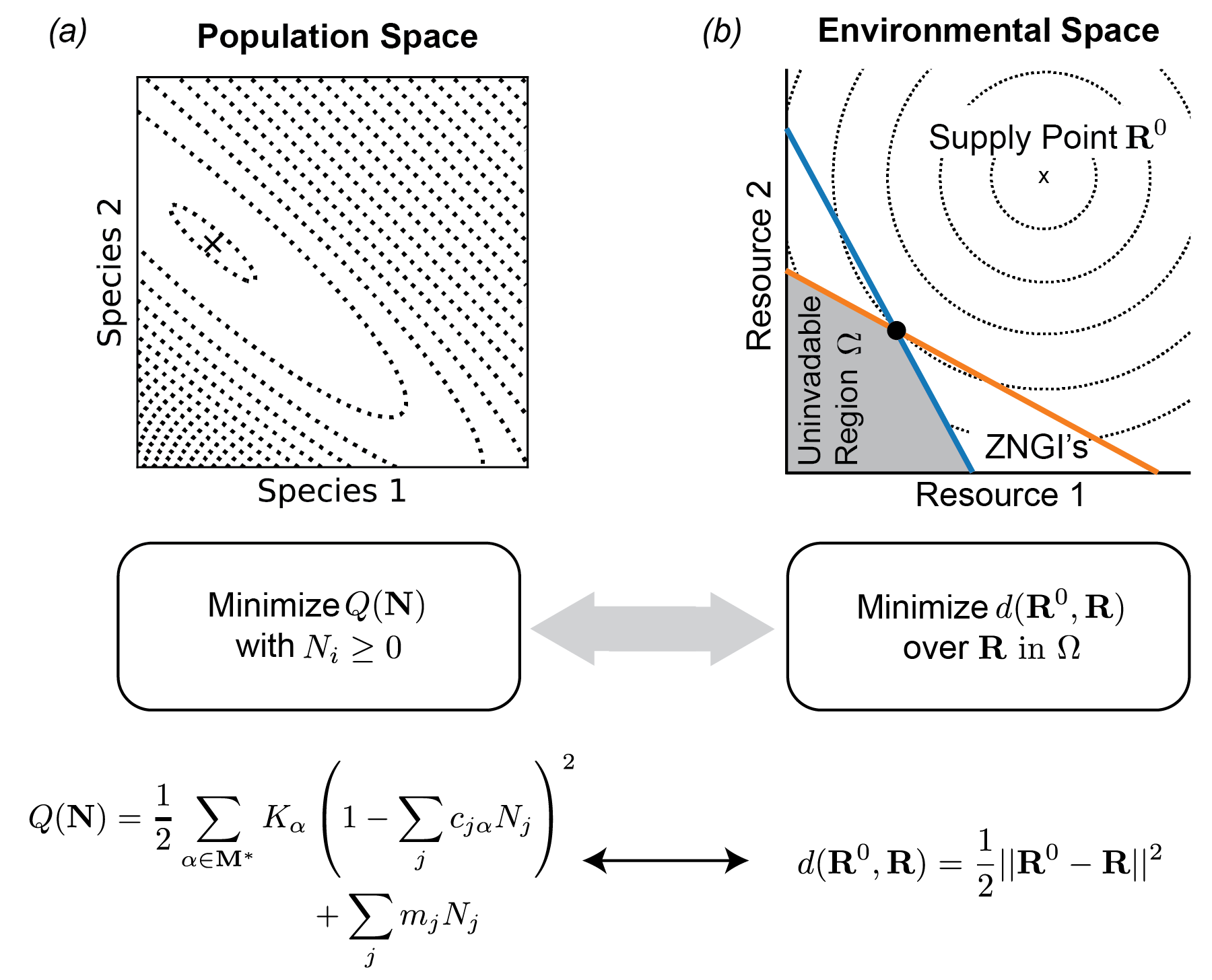}
	\caption{\linespread{1.3}\selectfont{} {\bf Reinterpreting MacArthur's Minimization Principle.} \emph{(a)} Contour lines of MacArthur's objective function $Q(\mathbf{N})$, in the space of population sizes, as defined in full in eq.~(\ref{eq:QN}). The `x' marks the equilibrium eventually attained in direct numerical simulation of eq.~(\ref{eq:MacN}-\ref{eq:MacR}) with the same parameters ($r_\alpha = m_i = w_\alpha = 1$ for $\alpha = 1,2$, $i = 1,2$; $K_1 = 4.8, K_2 = 2.85, c_{1\alpha} = (0.5,0.3), c_{2\alpha} = (0.4,0.6)$). The direct simulation ends up at the point where $Q$ is minimized, as predicted by MacArthur. Also shown for illustration is a simplified expression for $Q$ with the $r_\alpha$ and $w_\alpha$ set to 1. \emph{(b)} Contour lines in the environmental space of resource abundances, representing the dissimilarity measure $d(\mathbf{R}^0,\mathbf{R})$ with respect to the supply point $\mathbf{R}^0$. The uninvadable equilibrium state, indicated by the black dot, minimizes $d$ under the uninvadability constraint $g_i(\mathbf{R})\leq 0$, which constrains the environmental state to lie within the shaded region $\Omega$ bounded by the zero net-growth isoclines (ZNGI's, colored lines). For MacArthur's model of competition for noninteracting resources, with $r_\alpha = w_\alpha = 1$ as in the previous panel, $d$ is simply the Euclidean distance.}
	\label{fig:intro}
\end{figure*}

\section*{Background}
\label{sec:back}
\subsection*{MacArthur's Minimization Principle}
MacArthur originally considered a model of competition among $S$ consumer species for $M$ substitutable resources (\citealt{macarthur1970species}). The resources, with population densities $R_\alpha$ ($\alpha = 1,2,\dots M$) do not interact with each other directly, and each resource type is independently self-limiting with carrying capacity $K_\alpha$. The dynamics of the consumer population densities $N_i$ ($i = 1,2,\dots S$) and the resource abundances are described by the following set of differential equations:
\begin{align}
\frac{dN_i}{dt} &= e_i N_i \left[ \sum_{\alpha} w_\alpha c_{i\alpha} R_\alpha - m_i\right]\label{eq:MacN}\\
\frac{dR_\alpha}{dt} &= \frac{r_\alpha}{K_\alpha} R_\alpha (K_\alpha - R_\alpha) - \sum_i N_i c_{i\alpha} R_\alpha\label{eq:MacR}
\end{align}
where $c_{i\alpha}$ is the successful encounter rate of species $i$ searching for resource $\alpha$, $m_i$ is the ``maintenance cost'' or threshold consumption level for growth, $w_\alpha$ is the per-capita ``weight'' or nutritional value of each resource, $e_i$ is the quantity of nutritional value required for reproduction of a given species, and $r_\alpha$ is the low-density resource growth rate. A central feature of interest in any such model is the location of the stable equilibrium state $\bar{\mathbf{N}},\bar{\mathbf{R}}$. MacArthur showed that this state can be identified by eliminating the $R_\alpha$'s and minimizing a quadratic function of the $N_i$'s.

To eliminate the $R_\alpha$'s, MacArthur assumed that the resources relax quickly to the equilibrium state corresponding to the current consumer population sizes. Solving for $R_\alpha$ as a function of $N_i$ in the equilibrium equations $dR_\alpha/dt = 0$, one obtains a closed set of dynamics for the consumer population sizes:
\begin{align}
\frac{dN_i}{dt} &= e_i N_i\left[\sum_{\alpha \in \mathbf{M}^*} r_\alpha^{-1} K_\alpha w_\alpha c_{i\alpha} (r_\alpha -  \sum_j N_j c_{j\alpha})-m_i\right].
\end{align}
Here the set $\mathbf{M}^*$ is comprised of resources with feasible abundances $r_\alpha -  \sum_j N_j c_{j\alpha} \geq 0$. Any resources not satisfying this constraint are driven to extinction under the full dynamics. MacArthur noticed that these differential equations can be written in terms of the gradient of a quadratic function of the $N_i$'s:
\begin{align}
\label{eq:QN}
\frac{dN_i}{dt} &= - e_i N_i \frac{\partial Q}{\partial N_i}
\end{align}
with
\begin{align}
Q(\mathbf{N}) &= \frac{1}{2} \sum_{\alpha \in \mathbf{M}^*} r_\alpha^{-1} K_\alpha w_\alpha \left(r_\alpha - \sum_j c_{j\alpha} N_j \right)^2 + \sum_j m_j N_j,
\end{align}
as is easily verified by performing the partial derivative and comparing with the original equation. Equation (\ref{eq:QN}) implies that $\partial Q/\partial N_i = 0$ in equilibrium for all non-extinct populations $i$. The negative sign guarantees that this stationary point is a local minimum rather than a maximum. For the extinct populations, stability against re-invasion requires $\partial Q/\partial N_i > 0$. This means that setting $N_i = 0$ also minimizes $Q$ along these directions, subject to the feasibility constraint $N_i \geq 0$ (\citealt{gatto1982comments,gatto1990general}). We have plotted $Q(\mathbf{N})$ for a community with two consumer species in fig.~\ref{fig:intro}(\emph{a}), along with the equilibrium state eventually reached in a numerical simulation of eq.~(\ref{eq:MacN}-\ref{eq:MacR}). 

This result was an important step forward in understanding the nature of equilibrium states in this model. It shows, for example, that there is only one stable equilibrium state, since $Q$ is a convex function with a single local minimum. But this theorem as it stands is subject to several significant limitations. First of all, the restriction of the sum to the subset $\mathbf{M}^*$ of resources with $r_\alpha - \sum_j c_{j\alpha} N_j \geq 0$ makes the objective function more complicated than it initially seems, since it is actually a piecewise function consisting of sectors that are linear along some axes and quadratic along others. This seems not to have been noticed by MacArthur, who took the sum over all $M$ resources, or in subsequent discussion of his work (\citealt{macarthur1970species,Case1980,gatto1982comments,gatto1990general}). In fact, the restriction of the sum problematizes the ecological interpretation MacArthur achieved for one special case of the model, as discussed in Appendix A. Secondly, it remains unclear what assumptions are actually required to obtain a minimization principle. MacArthur took some steps in that direction by extending his result to the case of interacting resources and of competition to avoid predators. He noted that for all these cases, the key feature required was the symmetry of the interaction matrix in an effective Lotka-Volterra description of the scenario. But his approach cannot be straightforwardly applied to other important scenarios such as when abiotic nutrients are supplied by a chemostat, or when the growth kinetics saturate at high resource abundance.

\subsection*{Contemporary niche theory}
To address these limitations of MacArthur's result, we draw on the theoretical framework of contemporary niche theory, as consolidated by Chase and Leibold (\citealt{chase2003ecological}). We use this framework to generalize MacArthur's insight about the symmetry of the environmentally mediated interactions, obtaining a minimization principle valid for all niche models that are symmetric in the relevant sense defined precisely below.

\begin{table}
	\centering
	\begin{tabular}{|c|c|c|}
		\hline
		Symbol & Description & MCRM\\
		\hline
		$N_i$ & Species abundance & Consumer population density ([individuals][length]$^{-D}$) \\
		\hline
		$R_\alpha$ & Environmental factor & Resource population density ([individuals][length]$^{-D}$) \\
		\hline
		$g_i$ & Growth rate & $e_i \left[\sum_\alpha w_\alpha c_{i\alpha} R_\alpha-m_i\right]$ ([time]$^{-1}$)\\
		\hline
		$q_{i\alpha}$ & Impact vector & $-c_{i\alpha}R_\alpha$ ([length]$^D$[time]$^{-1}$)\\
		\hline
		$h_\alpha$ & Supply vector & $\frac{r_\alpha}{K_\alpha}(K_\alpha - R_\alpha)$ ([individuals][length]$^{-D}$[time]$^{-1}$)\\
		\hline
		$R_\alpha^0$ & Supply point & $K_\alpha$ ([individuals][length]$^{-D}$)\\
		\hline
	\end{tabular}
\caption{{\bf Key quantities of niche theory}. Final column lists how each quantity appears in MacArthur's Consumer Resource model (eq.~(\ref{eq:MacN}-\ref{eq:MacR})), along with its units. $D$ is the spatial dimension of the ecosystem (=2 for terrestrial, 3 for aquatic).}
\label{tab:niche}
\end{table}

Table \ref{tab:niche} lists the key elements of the theory, which aims to extract the essential features of MacArthur's Consumer Resource Model (MCRM, eq. \ref{eq:MacN}-\ref{eq:MacR} above). The first of these features is the explicit consideration of the environment, with the abundances $R_\alpha$ of the $M$ resources appearing alongside the population densities $N_i$ of the $S$ consumer species. Niche theory follows this basic scheme, but with a broader notion of ``resource'' that includes any environmental factor that affects an organism's growth rate (cf. \citealt{levin1970community}, \citealt{tilman1982resource}). In microbial ecology, for example, concentrations of quorum sensing molecules and antibiotics can act as resources in this extended sense (\citealt{momeni2017lotka}). 

The second feature of the MCRM is that the reproductive rates of the consumers depend only on the state of the environment, as specified by the resource abundances. Niche theory preserves this assumption, but allows this dependence to be described by an arbitrary set of functions $g_i(\mathbf{R})$. The consumers in the MCRM also affect the environment by depleting resources, with the per-capita depletion rate depending only on the resource abundances. In niche theory, this assumption is encoded by representing the impact of the organisms on their environment by a set of per-capita ``impact vectors,'' with the impact of species $i$ on resource $\alpha$ described by a function $q_{i\alpha}(\mathbf{R})$ (\citealt{tilman1982resource,leibold1995niche}). In the MCRM, the impacts are closely related to the growth rates, since resource contribute to the growth rate only insofar as they are removed from the environment. But generalized resources can affect the growth rate in other ways (e.g., production of antibiotics specifically inhibiting growth of other species), so the niche theory framework allows the impact vectors to be defined by an independent set of arbitrary functions.

Finally, the resources in the MCRM have their own intrinsic dynamics, described by a set of independent logistic growth laws. Niche theory places no constraints on the form of the intrinsic resource dynamics, which are described by a ``supply vector'' with elements $h_\alpha(\mathbf{R})$ (\citealt{tilman1982resource,chase2003ecological}). Generally, however, it is assumed that these dynamics have some stable equilibrium $\mathbf{R}_\alpha^0$, which is known as the ``supply point.'' 

These definitions lead to the following set of differential equations describing the population and environmental dynamics in a general niche model:
\begin{align}
\frac{dN_i}{dt} &= N_i g_i(\mathbf{R})\label{eq:niche1}\\
\frac{dR_\alpha}{dt} &= h_\alpha(\mathbf{R}) + \sum_i N_i q_{i\alpha}(\mathbf{R}).\label{eq:niche2}
\end{align}

\subsection*{Graphical analysis with ZNGI's}
The central assumption of niche theory is that all interactions between species are mediated by environmental factors, so that $g_i(\mathbf{R})$ and $q_{i\alpha}(\mathbf{R})$ are functions of the environmental state $\mathbf{R}$ alone, and are independent of the population sizes $N_i$. This assumption makes it possible to graphically analyze the equilibrium states of these models in resource space (\citealt{tilman1982resource}).  Central to the graphical approach is the hypersurface where $g_i(\mathbf{R})  = 0$, called the zero-net-growth isocline (ZNGI), depicted in a two-resource example in fig.~\ref{fig:intro}(\emph{b}) (\citealt{tilman1982resource,leibold1995niche}). Environmental states along the ZNGI support reproduction rates that exactly balance death rates, leading to constant population sizes. The ZNGI's play an essential role in the formulation of our new optimization principle.

For a given collection of species, the ZNGI's fix the boundaries of the ``uninvadable'' region $\Omega$ illustrated in fig.~\ref{fig:intro}(\emph{b}), defined as the set of environmental states $\mathbf{R}$ satisfying $g_i(\mathbf{R})\leq 0$ for all species $i$. All stable equilibrium states lie within this region, for any choice of supply vector and impact vector. The interior of the region does not support growth of any species in the collection, so any interior point can trivially be made into an ``empty'' stable equilibrium state by simply placing the supply point there and driving all the consumer species extinct. All non-empty equilibrium points lie on the boundary of $\Omega$, which is the outer envelope of the ZNGI's of all the species in the pool (cf. \citealt{koffel2016geometrical}). Points that lie on a ZNGI but are outside of $\Omega$ can be valid equilibrium states, but are unstable against invasion by species within the focal collection. 

\section*{Results}
\subsection*{General criteria for existence of optimization principle}
\label{sec:cond}
Our first main result is that MacArthur's observation on the conditions for the existence of an optimization principle can be extended to all models within the niche theory framework: \emph{equilibrium states of a niche-theory model optimize an objective function whenever the environmentally-mediated interactions among species are symmetric.}

Interaction symmetry is usually treated within the context of a generalized Lotka-Volterra model, which represents the interactions with a matrix of constant coefficients. But it can be defined more generally within niche theory by considering a small externally imposed perturbation in the abundance of a given species from some reference state. This perturbation will slightly shift the equilibrium resource abundances, which will in turn affect the growth rates of the other species. These environmentally mediated actions are symmetric if the effect on the growth rate of species $j$ of a change in the abundance of species $i$ is the same as the effect on species $i$ of the same change in species $j$. 

When all species are very similar to each other, this condition is straightforward to evaluate. But when species significantly differ in body size or other important characteristics, the quantification of abundance becomes ambiguous. In the case of body size differences, measuring the population in terms of total biomass gives a very different result than counting the number of individuals. This makes it unclear whether ``the same'' change in abundance is the same additional number of individuals or the same increase in total biomass. Whether or not the interactions are symmetric will depend on the choice of unit of measurement. 

To resolve this ambiguity, we define the interactions to be symmetric whenever there is at least one way of quantifying abundance under which symmetry is achieved. Mathematically, this can be expressed as the requirement that $\frac{dg_i}{d(a_j N_j)} = \frac{dg_j}{d(a_i N_i)}$ for some choice of positive scaling factors $a_i$. This flexibility in the relevant notion of symmetry was already noted by M. Gatto in the context of MacArthur's original work (\citealt{gatto1982comments}), and is here generalized to arbitrary models within the niche theory framework. It is actually slightly more general even than Gatto realized, because the scaling factors need not be constant, but can depend on the current state of the ecosystem. This flexibility makes a wide variety of resource competition models symmetric in the relevant sense.

Since the effect of a change in population size on the environment is determined the impact vector, while the effect of the change in environment on other species is determined by their growth rates, symmetry clearly requires the growth rates and impact vectors to be related in a special way. In Appendix B, we show that the required relationship takes the following form:
\begin{align}
q_{i\alpha}(\mathbf{R}) = - a_i(\mathbf{R}) b_{\alpha}(\mathbf{R}) \frac{\partial g_i}{\partial R_\alpha},
\label{eq:cond1}
\end{align}
where $a_i$ is the scaling factor introduced above, and $b_\alpha$ are functions of $\mathbf{R}$ that are the same for all species, but can vary from resource to resource. Since the scaling factors $a_i$ have already been defined to be positive, the new functions $b_\alpha$ should also be positive so that each species acts on the resource in a way that limits its own growth (cf. \citealt{tilman1982resource}). This proof of eq.~(\ref{eq:cond1}) as the condition for symmetry is somewhat technical, but once it is established, we can substitute this expression into eq.~(\ref{eq:niche1}-\ref{eq:niche2}) to obtain the following set of conditions for a stable equilibrium:
\begin{align}
{\text Steady \,\,populations \,\,} 0 &= a_i N_i g_i(\mathbf{R})\label{eq:kkt1}\\
{\text Steady \,\,environment\,\,} 0 &= \frac{h_\alpha(\mathbf{R})}{b_\alpha(\mathbf{R})} - \sum_i a_i N_i \frac{\partial g_i}{\partial R_\alpha}\\
{\text Noninvasibility\,\,} 0 &\geq g_i(\mathbf{R})\\
{\text Feasible \,\,populations\,\,} 0 &\leq a_i N_i. \label{eq:kkt4}
\end{align}
These are almost identical to the well-known Karush-Kuhn-Tucker (KKT) conditions for constrained optimization under the constraints $g_i \leq 0$, with the scaled population sizes $a_i N_i$ playing the role of the generalized Lagrange multipliers (also called KKT multipliers), and with $\frac{h_\alpha(\mathbf{R})}{b_\alpha(\mathbf{R})}$ taking the place of the negative gradient of the optimized function (\citealt{boyd2004convex,bertsekas1999nonlinear, bishop:2006:PRML}). These conditions generalize the theory of Lagrange multipliers to the case of inequality constraints, with the first equation setting the Lagrange multipliers $a_i N_i$ to zero for points below the constraint surface ($g_i < 0$), where the constraint has no effect. The KKT conditions were also employed by Gatto in his analysis of MacArthur's Minimization Principle, and also appear in a different context within optimal foraging theory (\citealt{gatto1982comments}, \citealt{tilman1982resource}). 

It turns out, as shown in Appendix B, that interaction symmetry also requires that $\frac{h_\alpha(\mathbf{R})}{b_\alpha(\mathbf{R})}$ can be written as the gradient of a function, which we will call $d(\mathbf{R})$:
\begin{align}
\frac{\partial d}{\partial R_\alpha} = -\frac{h_\alpha(\mathbf{R})}{b_\alpha(\mathbf{R})}. \label{eq:dddR}
\end{align}
The equilibrium conditions listed above thus guarantee that $d(\mathbf{R})$ is locally extremized over the uninvadable region $\Omega$ defined above, bounded by the outer envelope of the ZNGI's. Since the intrinsic dynamics of the environment push the state $\mathbf{R}$ along the direction of the supply vector $\mathbf{h}$, eq.~(\ref{eq:dddR}) implies that this extremum is in fact a minimum. 

This result generalizes MacArthur's Minimization Principle to all niche models with symmetric environmentally-mediated interactions. Stable equilibria of such models can be determined in four basic steps:
\begin{enumerate}
	\item Find $b_\alpha$ and $a_i$ by comparing the impact vectors with the derivative of the growth rates using eq.~(\ref{eq:cond1}).
	\item Compute $d$ from $b_\alpha$ and the supply vector using eq.~(\ref{eq:dddR}).
	\item Impose constraints $g_i \leq 0$, requiring that the environment lie in the uninvadable region.
	\item Minimize $d(\mathbf{R})$ under these constraints. The minimizing value is the equilibrium state $\bar{\mathbf{R}}$ of the environment, and the Lagrange multipliers that enforce the constraints are equal to $a_i \bar{N}_i$.
\end{enumerate}

This formulation of the minimization principle can also be extended to models without symmetry, although its practical implementation as a means of finding equilibrium state becomes more complicated. Any impact vector $q_{i\alpha}$ can always be decomposed into a sum of two terms, one of the form given by eq.~(\ref{eq:cond1}), and another that accounts for the rest of the impact. The minimization principle can be recovered by simply fixing this second term to its equilibrium value, and treating it as part of the supply vector $h_\alpha$. Even though the size of this correction to $h_\alpha$ cannot be determined until the equilibrium state is already known, it is still possible to find the equilibrium with an iterative approximation algorithm, described in Appendix B. The modified $h_\alpha$ also often has a clear ecological interpretation, as shown in the asymmetric examples below. 

\subsection*{Objective function measures environmental perturbation from surviving species}
\label{sec:mepp}
Our second result is that the quantity $d(\mathbf{R})$ has a natural and universal ecological interpretation. From eq.~(\ref{eq:dddR}), we see that the unconstrained minimum of $d$ lies at the supply point $\mathbf{R}^0$ where $\mathbf{h}(\mathbf{R}^0) = 0$. Since this equation only defines the minimized function $d(\mathbf{R})$ up to a constant offset, we are free to set its minimum value to be zero: $d(\mathbf{R}^0) = 0$. We now have a quantity that is always positive, and equals zero only when the environment is in its unperturbed equilibrium state. This makes $d(\mathbf{R})$ a natural way of quantifying the ``distance'' to the supply point. To indicate the fact that the function measures the size of the change from $\mathbf{R}^0$ to $\mathbf{R}$, from now on we will put both of these vectors as arguments and write $d(\mathbf{R}^0,\mathbf{R})$. 

In light of this interpretation of the objective function, we can state the Minimum Environmental Perturbation Principle (MEPP), valid for all symmetric niche models: \emph{Uninvadable equilibrium states minimize the perturbation of the environment away from the supply point, subject to the constraint that no species in the regional pool has a positive growth rate.}

In the following sections, we report the minimized function $d$ for seven commonly used ecological models that can be cast in the language of contemporary niche theory. Out of the infinite variety of possible ways of quantifying environmental changes, we will see that this function holds a privileged status, since it naturally reflects the importance of a given change for the ecological dynamics of the community. Full derivations of all results can be found in Appendix C.

\begin{figure*}
	\includegraphics[width=16cm]{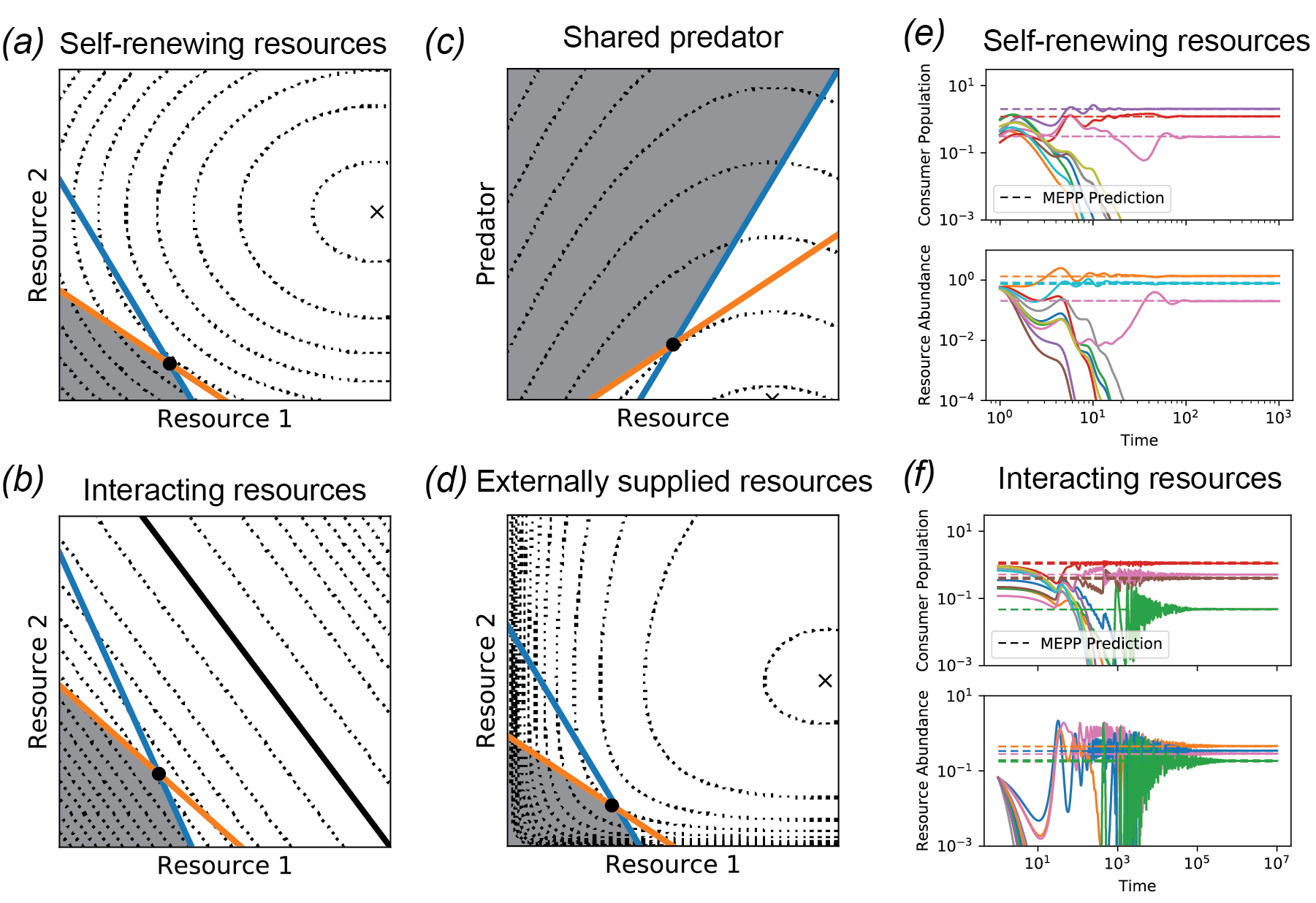}
	\caption{\linespread{1.3}\selectfont{} {\bf Examples with symmetric interactions.} \emph{(a,b,c,d)} ZNGI's, uninvadable region $\Omega$ and contours of perturbation measure $d(\mathbf{R}^0, \mathbf{R})$ for the four examples discussed in the text where the environmentally mediated interactions are symmetric, and MEPP straightforwardly applies. The black dot indicates the final state of a numerical simulation of the corresponding differential equations. \emph{(e,f)} Simulations of two of the models with larger numbers of species and resources, compared with the predictions of MEPP for the uninvadable equilibrium state. Consumer abundances are obtained from the Lagrange multipliers that enforce the constraints during optimization. See Appendix D or Jupyter notebooks for all simulation parameters. }
	\label{fig:symmetric}
\end{figure*}

\subsection*{Symmetric examples}
We begin with the three models considered by MacArthur in his original paper on the minimization principle (\citealt{macarthur1970species}): the model of competition for noninteracting resources discussed above and two generalizations. The first of these allows for the resources to compete directly with each other (e.g., plants competing for space or water), and the second includes competition among consumers to avoid shared predators. MacArthur obtained minimization principles in the space of population sizes for all these models, under the condition that the environmentally mediated interactions among consumer species remain symmetric. By performing the minimization in resource space, we obtain a unified physical interpretation in terms of the environmental perturbation, with the different models giving rise to different perturbation measures $d$, which reflect the ways in which environmental changes impact the community. This reinterpretation also allows us to readily generalize the minimization principle to a scenario not considered by MacArthur, where nutrients are supplied externally via a chemostat. 

Fig.~\ref{fig:symmetric} graphically depicts the optimization problem of each of these four scenarios, and also compares the results of constrained optimization of $d$ with direct numerical integration of the dynamical equations for two of them. See supplemental figure \ref{fig:extra} for simulations of the other two examples.

\subsubsection*{Noninteracting resources}
We begin with MacArthur's primary model, presented in eqs.~(\ref{eq:MacN}-\ref{eq:MacR}) above. This is a model of competition for non-interacting self-renewing resources, whose intrinsic population dynamics in the absence of consumers are described by independent logistic growth laws. The objective function in this case is simply the weighted Euclidean distance of the resource abundance vector from the supply point:
\begin{align}
d(\mathbf{R}^0,\mathbf{R}) &= \frac{1}{2}\sum_\alpha w_\alpha r_\alpha K_\alpha^{-1} (R_\alpha - R_\alpha^0)^2, \label{eq:dmcrm}
\end{align}
where the supply point $R_\alpha^0 = K_\alpha$ is here simply equal to the vector of resource carrying capacities. The contribution of each resource to this distance is weighted by the ecological significance of changes in its abundance. This weight has three components. The first factor, $w_\alpha$, measures the nutritional value of the resources. Resources with low values of $w_\alpha$ contribute less to the growth for consumer populations, and changes in their abundance are therefore less important. The second factor, $r_\alpha$, controls the rate of resource renewal. Abundances of resources with high rates of self-renewal are more difficult to perturb than those of resources that grow back slowly, and so a given shift in abundance is more significant for the former than for the latter. Finally, the factor of $K_\alpha^{-1}$ reflects the fact that a perturbation of the same absolute size is less significant if the carrying capacity is larger. 

As discussed in Appendix C, an important feature of the optimization perspective in all of MacArthur's examples is that the resource feasibility constraint $R_\alpha \geq 0$ must be enforced explicitly. This causes difficulties for the interpretation of MacArthur's original principle in the space of population sizes, but it fits easily into the resource space picture. For any niche model, the minimization is always subject to the constraint that the environment must lie in the uninvadable region $\Omega$, and the feasibility condition simply means that the lower boundary of this region must also be included in the optimization protocol.

\subsubsection*{Interacting self-regulation}
In reality, self-renewing resources like plants or algae usually compete directly with each other for space, water, light and nutrients. MacArthur therefore generalized his model to allow for this kind of interaction. For concreteness, we focus on the case where the resources are plants competing for space, with each individual of species $\alpha$ occupying an area $a_\alpha$. In this two-dimensional example, $N_i$ and $R_\alpha$ are both naturally measured in units of individuals per land area. The fraction of the land that is available for new plant growth is then given by $1-\sum_\alpha a_\alpha R_\alpha$. This results in the following set of equations, with the per-capita growth rate of the plants equal to a bare rate $r_\alpha$ times the free space fraction:
\begin{align}
\frac{dN_i}{dt} &= e_i N_i\left[ \sum_{\alpha} w_\alpha c_{i\alpha} R_\alpha - m_i\right]\label{eq:space1}\\
\frac{dR_\alpha}{dt} &= r_\alpha R_\alpha\left(1 - \sum_\beta a_\beta R_\beta \right) -  \sum_i c_{i\alpha} N_i R_\alpha.\label{eq:space2}
\end{align}
As MacArthur points out, the model with interacting resources requires additional assumptions to guarantee symmetry (\citealt{macarthur1970species}). In Appendix C, we show that a sufficient assumption is to make the growth rates $r_\alpha$ the same for all resources ($r_\alpha = r$), and the nutritional value of each plant species proportional to its size ($w_\alpha = w a_\alpha$). In this case, the objective function is 
\begin{align}
d(\mathbf{R}) &=  \frac{wr}{2}\left(1 - \sum_{\alpha} a_\alpha R_\alpha \right)^2.
\end{align}
We have dropped the $\mathbf{R}^0$ from the argument of $d$ for this example, because in the absence of consumers there is a multiplicity of equivalent unperturbed equilibrium states. In fact, every combination of plants that fills all the available space is a possible equilibrium. The objective function straightforwardly measures the perturbation away from this set of states, and is simply proportional to the square of the free area fraction.  

\subsubsection*{Competition to avoid predators}
MacArthur's final example adds another trophic level, allowing the consumer species to compete to avoid predators in addition to the competition for resources. The predators contribute an extra mortality term to the dynamics for the consumer population densities, which depends on the predator densities $P_a$ ($a = 1,2,3\dots M_P$). If we assume the same mass-action model for predation as for primary resource consumption, we obtain the following model:
\begin{align}
\frac{dN_i}{dt} &= e_i N_i \left[ \sum_{\alpha} w_\alpha c_{i\alpha} R_\alpha - m_i \right] - \sum_a p_{ia} P_a N_i\label{eq:pred1}\\
\frac{dR_\alpha}{dt} &= \frac{r_\alpha}{K_\alpha} R_\alpha (K_\alpha - R_\alpha) - \sum_i N_i c_{i\alpha} R_\alpha\\
\frac{dP_a}{dt} &= \sum_i \eta_i p_{ia} N_i P_a - u_a P_a\label{eq:pred3}
\end{align}
where $p_{ia}$ is the rate of predation of predator $a$ on species $i$, $u_a$ is the intrinsic mortality rate for predator species $a$, and $\eta_i$ is the nutritional value for predators of consumer (prey) species $i$. 

MacArthur claims that this model generically produces symmetric interactions, probably because he was not considering the role of the consumer nutritional content $\eta_i$ (\citealt{macarthur1970species}). For arbitrary $\eta_i$ and $e_i$ this turns out to be false, but we show in Appendix C that symmetry is restored if we assume that the biomass conversion efficiencies of the consumers are proportional to the inverse of their nutritional values ($e_i = e_0/\eta_i$). This assumption is in fact well-motivated on physical grounds, since $e_i^{-1}$ measures the amount of excess consumption required to produce a new individual of species $i$. If more consumption is required to produce an individual of a given species, then that individual should also hold more nutritional value for its predators. 


In this symmetric case, we can obtain a minimization principle by treating the predators as components of the environment. The objective function is:
\begin{align}
d(\mathbf{R}^0,\mathbf{P}^0,\mathbf{R},\mathbf{P}) &= \frac{1}{2}\sum_\alpha \frac{r_\alpha}{K_\alpha} w_\alpha (R_\alpha - R_\alpha^0)^2 + \frac{1}{e_0} \sum_a u_a P_a
\end{align}
with supply point $R^0_\alpha = K_\alpha$, $P^0_a = 0$. This is the same as for the original consumer resource model, with the addition of the predator-dependent term $\sum_a u_a P_a$. The new term is minimized when all the predators are extinct, which is the ``unperturbed state'' for predators that cannot survive in the absence of prey. Each predator is weighted by its mortality rate, reflecting the same logic as the presence of $r_\alpha$ in weights of the resource perturbations. Finally, the balance between the importance of the resource and predator terms is set by $e_0$, which controls the efficiency of energy transfer between trophic levels. Perfect efficiency corresponds to $e_0 = 1$. Larger values of $e_0$ correspond to lower efficiency, which makes the contributions of the predators less important.

\subsubsection*{Externally supplied resources}
In the three preceding examples, resources are self-renewing with exponential growth at low densities. Microscopic ecosystems, however, are commonly maintained in the laboratory using serial dilutions, whereby a fraction $f$ of the sample volume is periodically transfered to fresh media with resource abundances $R_\alpha^0$ at time interval $T$, with the rest discarded or frozen for later analysis. This creates a new timescale $\tau = \frac{T}{1-f}$ over which the resource concentrations relax towards $R_\alpha^0$ in the absence of reproduction or consumption, leading to the following set of dynamical equations:
\begin{align}
\frac{dN_i}{dt} &= e_i N_i \left[ \sum_{\alpha} w_\alpha c_{i\alpha} R_\alpha-m_i\right] - \tau^{-1}N_i\label{eq:chemN}\\
\frac{dR_\alpha}{dt} &= \tau^{-1} (R_\alpha^0 - R_\alpha) - \sum_i N_i c_{i\alpha} R_\alpha. \label{eq:chemR}
\end{align}
Note that we have also added an extra term $\tau^{-1} N_i$ to the dynamics of the consumers, to account for the dilution of the consumer populations caused by this protocol. Adding this term is equivalent to modifying the maintenance cost $m_i$, but writing it explicitly allows us to preserve the physiological meaning of $m_i$ as an intrinsic property of the consumer species.

This model produces symmetric interactions between consumer species, regardless of the choice of parameter values. The objective function is no longer quadratic, however, but is given by a weighted Kullback-Leibler divergence:
\begin{align}
d(\mathbf{R}^0,\mathbf{R}) &= \tau^{-1} \sum_\alpha w_\alpha\left[R_\alpha^0 \ln \frac{R_\alpha^0}{R_\alpha} - (R_\alpha^0 - R_\alpha)\right]. \label{eq:KL}
\end{align}
This is a natural way of quantifying the difference between two vectors with all positive components, such as probabilities or chemical concentrations (\citealt{rao2016nonequilibrium}). As in the original MacArthur model, the contribution of each resource is weighted by its nutritional value $w_\alpha$. But now the feasibility constraint $R_\alpha\geq 0$ need not be enforced explicitly, because $d(\mathbf{R}^0,\mathbf{R})$ diverges as $R_\alpha \to 0$, guaranteeing that the constrained optimum will always lie in the feasible region.

\begin{figure*}
	\includegraphics[width=16cm]{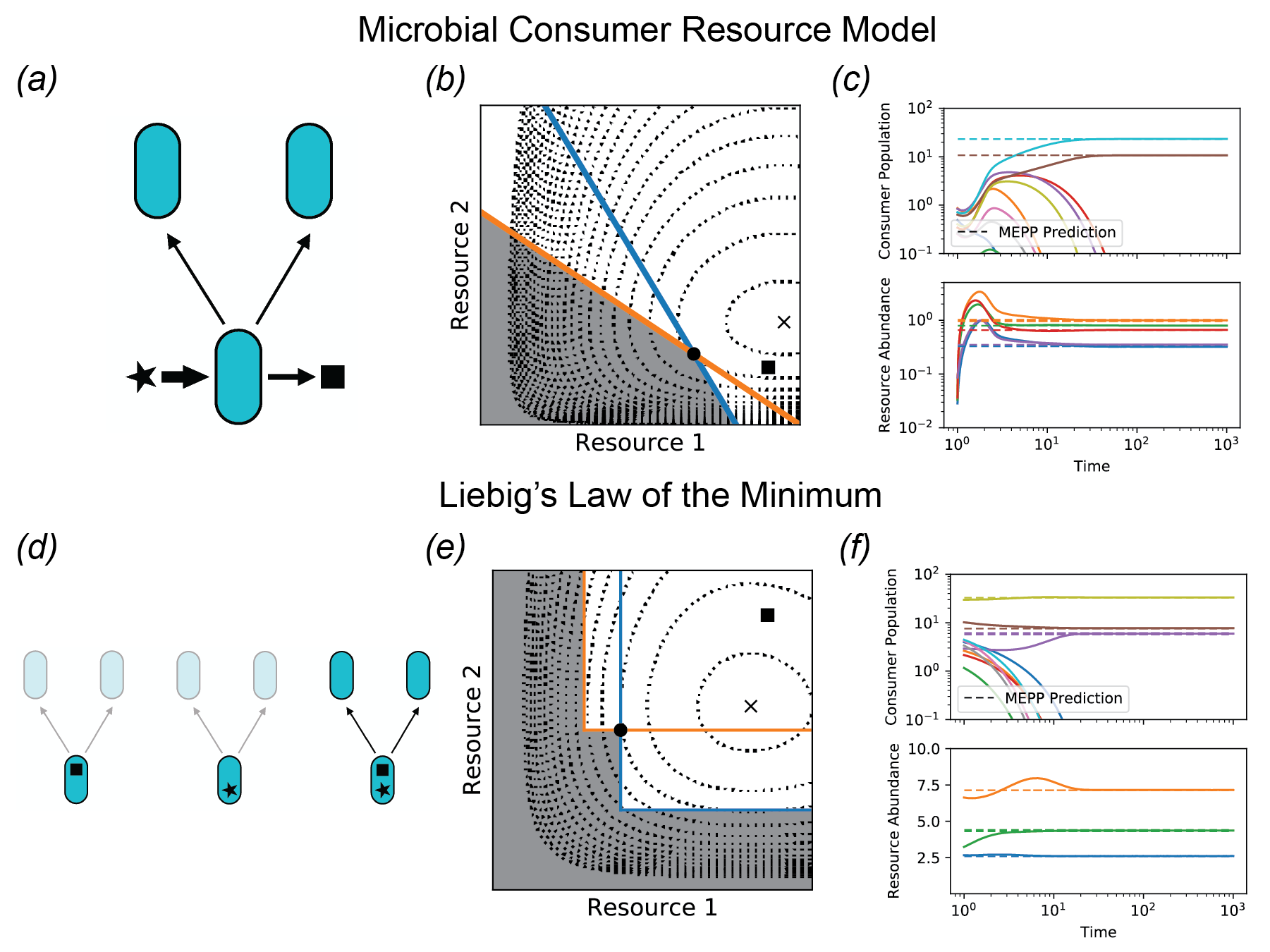}
	\caption{\linespread{1.3}\selectfont{} {\bf Examples with asymmetric interactions.} \emph{(a)} The Microbial Consumer Resource Model (MiCRM, eq.~\ref{eq:Mi1}-\ref{eq:Mi2}) describes microbial ecosystems where byproducts of resource metabolism can be used as growth substrates for other organisms. \emph{(b)} ZNGI's (colored lines) and the uninvadable region (shaded) for a pair of microbial species in the presence of two interconvertible  resources. Contour lines represent the function $d(\tilde{\mathbf{R}}^0,\mathbf{R})$ that is minimized in the uninvadable equilibrium state. Square is true supply point $\mathbf{R}^0$ and `x' is effective supply point $\tilde{\mathbf{R}}^0$ accounting for the byproducts generated in one chemostat turnover time $\tau$. Black dot is the equilibrium state reached by a direct numerical simulation.  \emph{(c)} Simulation of 10 microbial species and 5 resource types, along with extended MEPP predictions obtained using the iterative algorithm described in Appendix B. \emph{(d)} Liebig's Law of the Minimum (eq.~\ref{eq:lie1}-\ref{eq:lie2}) describes the dependence of an organism's growth rate on several essential nutrients (square and star), which must all be present in sufficient abundance in order for the organism to reproduce. \emph{(e)} ZNGI's, uninvadable region, objective function and supply points. \emph{(f)} Simulation of 10 species competing for 3 essential resources, along with extended MEPP predictions. See Appendix D or Jupyter Notebook for parameters.}
		\label{fig:asymmetric}
	\end{figure*}

\subsection*{Asymmetric examples}
We now turn to two important scenarios where interactions are unavoidably asymmetric: a recently introduced Microbial Consumer Resource Model where consumers generically produce metabolic byproducts, and competition for essential resources described by Liebig's Law of the Minimum. This fundamental asymmetry of these models results from the fact that organisms can affect the environment in ways that are unrelated to their own growth rate, whether by producing novel byproducts, or by consuming resource types that do not limit their growth. In this section, we describe how this ``extra'' supply or consumption is accounted for by a shift in the effective supply point $\tilde{\mathbf{R}}^0$, as illustrated in fig.~\ref{fig:asymmetric}. 

\subsubsection*{Microbial Consumer Resource Model}
The Microbial Consumer Resource Model (MiCRM) describes microbial consumers that generically produce metabolic byproducts, as illustrated in fig.~\ref{fig:asymmetric}\emph{(a)} (\citealt{Goldford2018,marsland2018available,marsland2019community}). A fraction $l_\alpha$ of the growth value resulting from uptake of resource $\alpha$ is released back into the environment, after being transformed into a variety of other resource types through internal metabolic reactions. A matrix $D_{\beta\alpha}$ specifies the fraction of byproduct from consumption of resource $\alpha$ that is released as resource $\beta$. This results in the following dynamical equations:
\begin{align}
\frac{dN_i}{dt} &= e_i N_i \left[\sum_\alpha (1-l_\alpha) w_\alpha c_{i\alpha} R_\alpha - m_i\right] \label{eq:Mi1}\\
\frac{dR_\alpha}{dt} &= \tau^{-1} (R_\alpha^0 - R_\alpha) - \sum_{i} N_i c_{i\alpha} R_\alpha + \sum_{i\beta}N_i D_{\alpha\beta}l_\beta\frac{w_\beta}{w_\alpha} c_{i\beta}R_\beta\label{eq:Mi2}.
\end{align}
The addition of byproduct secretion breaks the symmetry of the effective interactions in the original consumer resource model. When species A produces a byproduct that benefits species B, species B may not produce any byproduct accessible to species A. Even in cases where the exchange is mutual, there is no reason why the size of the benefit would be identical in both directions.

The equilibrium state of this model minimizes the same objective function as the chemostat model discussed above in eq.~(\ref{eq:KL}), but with a modified supply point
\begin{align}
\tilde{R}^0_\alpha = R^0_\alpha + \tau \sum_{i\beta}\bar{N}_i D_{\alpha\beta}l_\beta\frac{w_\beta}{w_\alpha} c_{i\beta}\bar{R}_\beta.
\end{align}
The second term is equal to the total quantity of resource $\alpha$ produced by all consumer species over the chemostat turnover time $\tau$. This modification thus accounts in an intuitive way for the extra supply due to byproduct secretion.  Fig.~\ref{fig:asymmetric} shows the location of the true supply point $\mathbf{R}^0$ and the effective supply point $\tilde{\mathbf{R}}^0$ for an example with two species and two resource types. Although the environment is directly supplied with very low levels of resource 2, the byproduct secretion moves the supply point up higher in that direction, allowing both species to coexist.  

As noted above, the correction to the supply point depends on the equilibrium population sizes $\bar{N}_i$ and resource abundances $\bar{R}_\beta$, and can therefore be calculated exactly only when the problem is already solved. But fig.~\ref{fig:asymmetric}\emph{(c)} shows that a simple iterative algorithm (described in Appendix B) successfully finds a self-consistent solution that agrees with direct numerical simulation.

\subsubsection*{Liebig's Law of the Minimum}
In all the examples presented above, resources were perfectly substitutable. But there are many ecological scenarios where different resource types serve different biological needs, and all of them must be simultaneously present at sufficient abundance in order to sustain growth, as illustrated in fig.~\ref{fig:asymmetric}\emph{(d)}. One typical example is competition of plants for nitrogen, phosporous and water, which are all required for the production of biomass. Such growth kinetics are commonly described by Liebig's Law of the Minimum, where the growth rate is determined by the availability of the most limiting resource. The standard choice of impact vector for this model assigns each species $i$ a constant stoichiometry $\nu_{\alpha i}$ which specifies the fraction of total consumption allocated to each resource type (\citealt{tilman1982resource,letten2017linking}). Using Michaelis-Menten growth kinetics for each resource, with maximum velocities $\mu_{i\alpha}$ and Michaelis constants $k_{i\alpha}$, we have
\begin{align}
\frac{dN_i}{dt} &= N_i \left[\underset{\beta}{\rm min}\left(\left\{\frac{\mu_{i\beta} R_\beta}{k_{i\beta}+R_\beta}\right\}\right) - m_i \right]\label{eq:lie1}\\
\frac{dR_\alpha}{dt} &= \tau^{-1} (R_\alpha^0 - R_\alpha) - \sum_{i} N_i \nu_{\alpha i} \underset{\beta}{\rm min}\left(\left\{\frac{\mu_{i\beta} R_\beta}{k_{i\beta}+R_\beta}\right\}\right)\label{eq:lie2}.
\end{align}
If each species were to deplete only its limiting resource, the effective interactions in this model would remain symmetric, and the equilibrium state would minimize the perturbation away from the true supply point, as measured by eq.~(\ref{eq:KL}). But this is biologically unreasonable, since the whole point of ``essential'' resources is that all of them must be taken up together in order to generate growth. The consumption of non-limiting resources shifts the effective supply point $\tilde{R}^0_\alpha$ by subtracting off the amount of each resource $\alpha$ consumed over the chemostat turnover time $\tau$ by organisms that are not limited by this resource. Fig.~\ref{fig:asymmetric}\emph{(e)} shows this drop in the supply point for an example with two resources and two consumers.

In fig.~\ref{fig:asymmetric}\emph{(f)}, we apply the same iterative scheme mentioned above to self-consistently obtain the equilibrium state and effective supply point, and compare the results with direct numerical simulation. This model generically exhibits multiple alternative stable states, and so care must be taken to ensure that both methods end up in the same one. In the simulations shown here, we simply initialized the direct simulation close to the MEPP prediction.

\begin{figure*}
	\includegraphics[width=16cm]{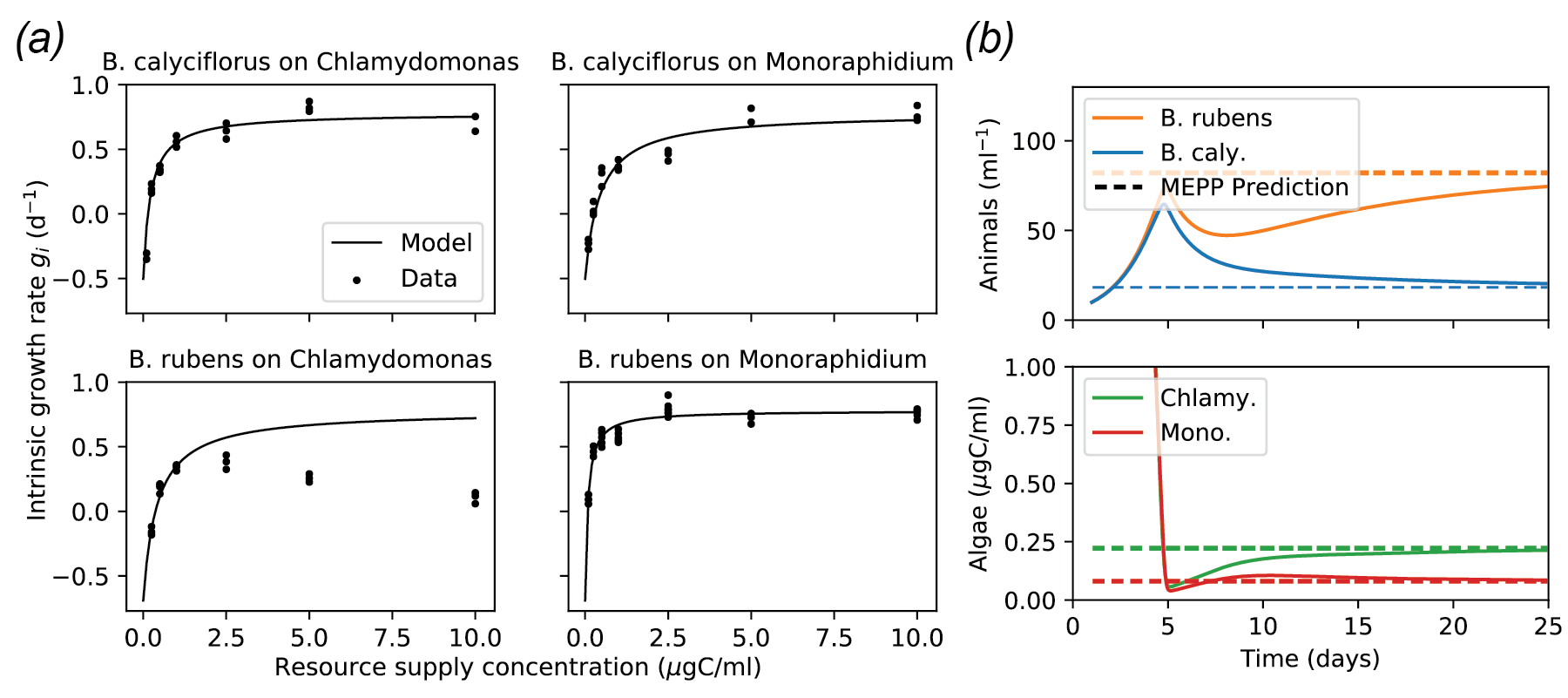}
	\caption{\linespread{1.3}\selectfont{} {\bf Applying MEPP to laboratory experiments.} \emph{(a)} Growth rate measurements reported in \citealt{rothhaupt1988mechanistic} for two species of zooplankton (\emph{Brachionus Rubens} and \emph{Brachionus calyciflorus}) fed with different concentrations of two species of algae (\emph{Chlamydomonas sphaeroides} and \emph{Monoraphidium minutum}). Black lines are simultaneous fits to eq.~(\ref{eq:II}) with resource-independent maximal uptake rates $J_{i\alpha} = J_i$, and with the maximal clearance rates $c_{i\alpha}$ equal to the directly measured values listed in table \ref{tab:roth}. Inferred parameter values are also listed in the table. For \emph{B. rubens} on \emph{Chlamydomonas}, additional ecological mechanisms came into play at high food densities that are not captured by a Type II growth model, and so only the three lowest densities were used for fitting. \emph{(b)} Simulations and MEPP predictions using the parameters in table \ref{tab:roth}, with supply point $w_c R_c^0 = 6$, $w_m R_m^0 = 4$ ($\mu$gC/ml) and $\tau = 5$ days.}
	\label{fig:roth1}
\end{figure*}
\subsection*{Application to zooplankton competition experiments}
In the 1980's, K. Rothhaupt performed a set of detailed experiments on resource competition in zooplankton to test Tilman's recent graphical formulation of niche theory (\citealt{rothhaupt1988mechanistic}). This study provides a convenient setting for illustrating how the key assumption of symmetric interactions can be confirmed or rejected, and how the perturbation $d(\mathbf{R}^0,\mathbf{R})$ can be measured.

 Fig.~\ref{fig:roth1} shows growth rates for the zooplankton \emph{Brachionus Rubens} and \emph{Brachionus calyciflorus} fed with different concentrations of the algae \emph{Chlamydomonas sphaeroides} and \emph{Monoraphidium minutum}. These plots show that the growth rates saturate at high levels of resource concentration, and it turns out that the relevant concentrations for the competition experiments lie well outside of the initial linear regime. This means we must consider a model that goes beyond any of the examples discussed above, and explicitly incorporates the saturation.

\subsubsection*{Saturating growth kinetics}
We model the saturation of the growth kinetics using Holling's Type II functional response, combining the contributions of the two resources in the manner appropriate to a well-mixed environment (\citealt{holling1959some,vincent1996trade}). We can use the same parameters with the same definitions as in the MCRM, but with the addition of a set of handling times $t_{i\alpha}$ for each consumer-resource pair:
\begin{align}
\frac{dN_i}{dt} &= e_i N_i \left[\sum_\alpha w_\alpha\frac{c_{i\alpha}R_\alpha}{1 + \sum_\beta c_{i\beta}t_{i\beta}R_\beta} - m_i \right] \label{eq:II}.
\end{align}
In the experiments of interest, the intrinsic value $w_\alpha$ of each species of algae is taken to be proportional to its carbon content, accounting for the significant difference in size between the two species, and resource abundances are reported as carbon concentrations $w_\alpha R_\alpha$. It is therefore convenient to analyze the model in terms of the maximum carbon uptake rate defined by
\begin{align}
J_{i\alpha} = \frac{w_\alpha}{t_{i\alpha}}
\end{align}
instead of using the handling time directly. 

In the wild, we would expect the resource equation for this system to have the same logistic supply vector as MacArthur's original model (eq.~\ref{eq:MacR}), with modified impact vectors to account for the saturation. But Rothhaupt's competition experiments follow the serial dilution protocol described above in the ``Externally supplied resources'' section, with algae supplied at a given concentration from an external source at fixed time intervals, and with experiments performed in the dark to minimize algae growth. We therefore use the chemostat supply vector of eq.~(\ref{eq:chemR}), and obtain
\begin{align}
\frac{dR_\alpha}{dt} &= \tau^{-1} (R_\alpha^0 - R_\alpha) - \sum_i N_i\frac{c_{i\alpha}R_\alpha}{1+\sum_\beta  c_{i\beta}t_{i\beta}R_\beta}. \label{eq:IIR}
\end{align}
In general, this model gives rise to asymmetric interactions. But they become symmetric when the maximum carbon uptake rates $J_{i\alpha}$ for each species $i$ are independent of the food source $\alpha$. In this case, as shown in Appendix C, MEPP applies and the equilibrium state minimizes the same objective function as the ordinary chemostat model given in eq.~(\ref{eq:KL}). The only difference is in the formula for the boundaries of the uninvadable region $\Omega$, which are now given by eq.~(\ref{eq:II}). In terms of the weighted concentration $w_c R_c$ of \emph{Chlamydomonas} and the concentration $w_m R_m$ of \emph{Monoraphidium}, with supplied concentrations $w_c R_c^0$ and $w_m R_m^0$, the perturbation measure is:
\begin{align}
d(\mathbf{R}^0,\mathbf{R}) &= \tau^{-1} \left[w_c R_c^0 \ln \frac{w_c R_c^0}{w_c R_c} + w_m R_m^0 \ln \frac{w_m R_m^0}{w_m R_m} - (wR^0-wR)\right] \label{eq:KLr}
\end{align}
where $wR = w_c R_c + w_m R_m$ and $wR^0 = w_c R_c^0 + w_m R_m^0$ are the total carbon concentrations in the ecosystem and in the supply, respectively. MEPP predicts that the equilibrium concentrations of \emph{Chlamydomonas} and \emph{Monoraphidium} minimize this function, subject to the constraint that the growth rates of both zooplankton species given by eq.~(\ref{eq:II}) are zero or negative.

\begin{table}
	\centering
	\begin{tabular}{|c|c|c|}
		\hline
		\multicolumn{3}{|c|}{\bf \emph{B. calyciflorus} parameters}\\
		\hline
		Symbol & Description & Value\\
		\hline
		$e_c$ & Individuals produced per unit carbon uptake & 13.8 /$\mu$g carbon \\
		\hline
		$J_c$ & Maximal carbon uptake rate & 0.0927 $\mu$g carbon/day\\
		\hline
		$m_c$ & Minimum viable carbon uptake rate & 0.0363 $\mu$g carbon/day\\
		\hline
		$c_{cc}$ & Maximal \emph{Chlamydomonas} clearance rate & 0.427 ml/day\\
		\hline
		$c_{cm}$ & Maximal \emph{Monoraphidium} clearance rate & 0.211 ml/day\\
		\hline
		\hline
		\multicolumn{3}{|c|}{\bf \emph{B. rubens} parameters}\\
		\hline
		Symbol & Description & Value\\
		\hline
		$e_r$ & Individuals produced per unit carbon uptake & 72.8 /$\mu$g carbon \\
		\hline
		$J_r$ & Maximal carbon uptake rate & 0.0202 $\mu$g carbon/day\\
		\hline
		$m_r$ & Minimum viable carbon uptake rate & 0.00947 $\mu$g carbon/day\\
		\hline
		$c_{rc}$ & Maximal \emph{Chlamydomonas} clearance rate & 0.0490 ml/day\\
		\hline
		$c_{rm}$ & Maximal \emph{Monoraphidium} clearance rate & 0.252 ml/day\\
		\hline
	\end{tabular}
	\caption{{\bf Parameter values for zooplankton competition experiments}. Maximal clearance rates are reproduced from a table
		 of measurements using radiolabeled algae reported in \citealt{rothhaupt1988mechanistic}, converted to a consistent set of units. The other parameters come from fitting eq.~(\ref{eq:II}) to an independent set of growth rate measurements reported in the same study, as shown in fig.~\ref{fig:roth1}.}
	\label{tab:roth}
\end{table}
\subsubsection*{Testing the model}
The key assumption about the maximum carbon uptake rates can be directly tested in principle by supplying the animals with large concentrations of each type of food, and checking whether the growth rates are the same in both cases. Fig.~\ref{fig:roth1} confirms that the maximum growth rate of \emph{B. calyciflorus} is indeed the same for both food sources, to within experimental uncertainty. The growth kinetics of \emph{B. rubens} at large \emph{Chlamydomonas} concentrations are non-monotonic, however, which Rothhaupt attributes to mechanical disturbance of the feeding process that is not reflected in Holling's Type II growth law (\citealt{rothhaupt1988mechanistic}). Thus we can only use the model in eq.~(\ref{eq:II},\ref{eq:IIR}) for this case at low food concentrations, where this additional mechanism can be neglected. The resource-independence of $w_\alpha/t_{i\alpha}$ can therefore only be tested indirectly for this organism, using the goodness of fit of the low-concentration data points to eq.~(\ref{eq:IIR}) when this condition is imposed.

There is also a second, hidden assumption, which was already made in MacArthur's original model with linear growth kinetics, concerning the dual role of the parameter $c_{i\alpha}$. This parameter has units of volume/time in an aquatic scenario, and mechanistically represents a ``clearance rate,'' that is, the volume of water cleared of food organisms by an individual consumer per unit time. In the saturating model the actual clearance rate is a function of food density, but $c_{i\alpha}$ still represents the maximal clearance rate, when food is scarce and handling time is not the limiting factor. This parameter can thus be directly measured by simply counting the number of food organisms ingested by an individual consumer over a short period of time over which the food density is approximately constant. Rotthaupt carried out such measurements using radiolabeled algae, and reported the maximum clearance rates for all four consumer-resource pairs. The mean values over at least 10 independent measurements are reproduced in table \ref{tab:roth} (see \cite{rothhaupt1988mechanistic} for complete methods, number of replicates and uncertainties). 

The assumption made in both the MCRM and in eq.~(\ref{eq:II},\ref{eq:IIR}) above is that the same parameters $c_{i\alpha}$ also determine the relative effects of different resource types on the consumer growth rate. To test this assumption, we performed a simultaneous nonlinear regression of eq.~(\ref{eq:II}) for each species to sets of growth rate measurements on both food sources, as shown in fig.~\ref{fig:roth1}. The clearance rates $c_{i\alpha}$ were held fixed at their directly measured values, and the maximum carbon uptake rates were assumed to be independent of food source, leaving three free parameters $e_i$, $J_i$ and $m_i$. The best-fit values are tabulated along with the clearance rates in table \ref{tab:roth}. These three parameters are sufficient to provide an excellent fit to both growth curves, with the exception of the high \emph{Chlamydomonas} concentrations with \emph{B. rubens} mentioned above. 

\begin{figure*}
	\includegraphics[width=16cm]{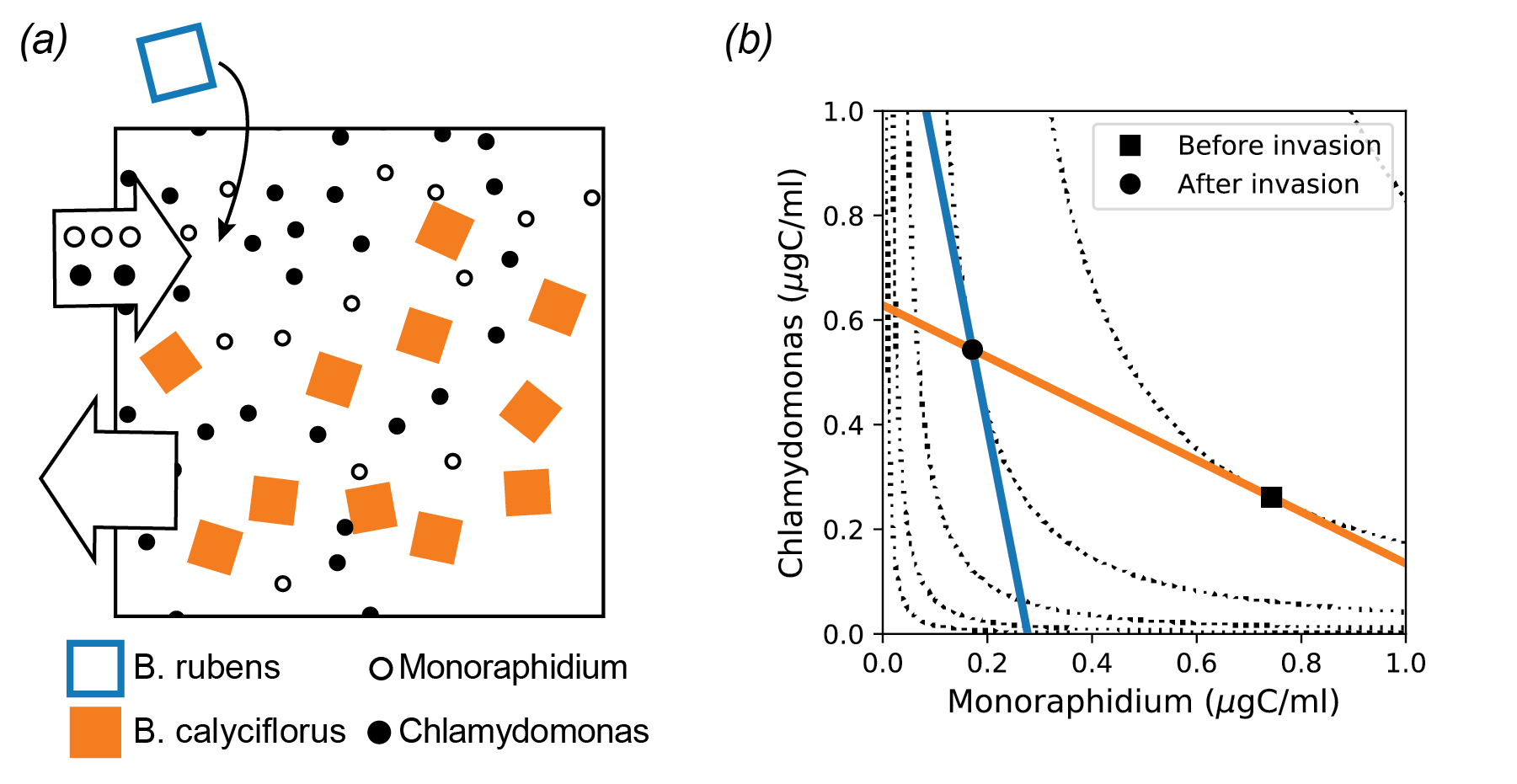}
	\caption{\linespread{1.3}\selectfont{} {\bf Consequences for community assembly.} \emph{(a)} Schematic of a hypothetical community assembly experiment, taking place in a chemostat supplied with a constant influx of \emph{Monoraphidium} and \emph{Chlamydomonas} as food. The system is first allowed to relax to equilibrium with \emph{B. calyciflorus} as the only consumer species, and then \emph{B. rubens} is added to the chamber. \emph{(b)} ZNGI's (solid colored lines) and contour lines of $d$ (dotted) using the experimentally determined parameters in table \ref{tab:roth}, at dilution rate $\tau^{-1} = 0.45$/day and supply levels $w_c R_c^0 = 6$, $w_m R_m^0 = 4$ ($\mu$gC/ml). The initial equilibrium of the assembly experiment is indicated by the black square, where $d$ is minimized subject only to the constraint that \emph{B. calyciflorus} has a vanishing growth rate. The final equilibrium is represented by the black circle, and lies on a higher contour line of $d$.}
	\label{fig:roth2}
\end{figure*}
\subsection*{Consequences for community assembly and eco-evolution}
In addition to providing a clear interpretation of MacArthur's principle and facilitating generalization, MEPP makes new predictions for scenarios where new species are added to an existing community. Specifically, MEPP implies that the perturbation measure $d(\mathbf{R}^0,\mathbf{R})$ is a monotonically increasing function under successive invasions, for any monostable niche model with symmetric environmentally-mediated interactions. 

To illustrate this result, we consider the hypothetical community assembly experiment depicted in fig.~\ref{fig:roth2}. We take the setup of K. Rothhaupt described above, and start with both resources present but with \emph{B. calyciflorus} as the only consumer. We perform serial dilutions according to the same protocol until species and resource abundances reach equilibrium, and then invade with \emph{B. rubens}. MEPP implies that the initial equilibrium minimizes $d(\mathbf{R}^0,\mathbf{R})$ of eq.~(\ref{eq:KLr}) under the single constraint that the net growth rate of \emph{B. calyciflorus} vanishes. When \emph{B. rubens} is introduced, a second constraint is added, leading to a new constrained optimum $\bar{\mathbf{R}}$. Since the new optimization is subject to more constraints, the new minimum is necessarily further from zero than the original, as is clear from the figure. 

The fact that $d$ monotonically increases under all successful invasions has a number of significant consequences. First of all, it implies that community assembly and evolution are unidirectional processes, just as one na\"{i}vely expects, and that limit cycles or chaos in the space of community compositions is ruled out (cf. \citealt{doebeli2017point}). In fact, given two snapshots of an evolving system, one can determine which came earlier and which came later by measuring the resource abundances and computing $d$. Without knowing anything about the consumer species, we can say that the snapshot with the higher value of $d$ must have come later. This also makes it possible to rule out possible trajectories for community assembly. If one observes two systems with the same resource supply in different equilibrium states, one can determine whether one of them can be assembled from the other by invading with the missing species. If community A has a larger value of $d$ than community B, then changing the composition of A to match B requires directly killing off some species, and cannot be accomplished through any set of successive invasions.

\section*{Discussion}
The Minimum Environmental Perturbation Principle provides a new perspective on niche theory, which opens up a number of interesting avenues for further investigation. First of all, measuring the environmental perturbation $d$ could shed light on the robust empirical correlation between diversity and productivity (\citealt{tilman2014biodiversity}). Since each species places an independent constraint on the domain of optimization, as noted above, $d$ will be positively correlated with species richness whenever MEPP applies. Larger $d$ means that the equilibrium resource abundances are further from the supply point for more diverse communities, which typically implies that more of the available resources are being converted to biomass. In cases where increased diversity fails to improve biomass yields, part of the explanation may lie in a significant asymmetry in the interactions that causes a major shift in the effective supply point.

MEPP also has important implications for evolution. It was recently shown that the graphical methods of niche theory can be applied to evolution through consideration of a continuum of ZNGI's, representing all possible phenotypes (\citealt{koffel2016geometrical}). Any evolutionarily stable phenotype (or collection of coexisting phenotypes) must lie on the outer envelope formed by all these ZNGI's. Since MEPP is valid for any number of species, it also applies to this continuum limit as long as the essential condition of interaction symmetry holds. 

The fact that $d$ is strictly non-decreasing under sequential invasions also suggests a connection to recent work on evolutionary optimization in the presence of environmental feedbacks (\citealt{metz2008when}). By computing the minimum value of $d$ for every possible combination of coexisting phenotypes, one can construct a community-level ``fitness landscape'' on which all evolutionary trajectories always travel monotonically uphill. These connections have yet to be fully explored, and remain an important area for future study.

\section*{Acknowledgments}
This work was supported by NIH NIGMS grant 1R35GM119461 and Simons Investigator in the Mathematical Modeling of Living Systems (MMLS) to PM. We would also like to thank Ching-Hao Wang,  Jacob Ferguson and William Ludington for useful discussions.

\bibliographystyle{amnatnat}
\bibliography{../literature/references.bib}
\newpage{}
\renewcommand{\theequation}{A\arabic{equation}}
\renewcommand{\thetable}{A\arabic{table}}
\renewcommand{\thefigure}{A\arabic{figure}}
\setcounter{equation}{0}  
\setcounter{figure}{0}
\setcounter{table}{0}
\section*{Appendix A: Interpretations of MacArthur's Minimization Principle}
MacArthur developed an interpretation of the minimization principle for the consumer resource model with non-interacting resources, under the assumption that all species have the same requirements $m_i = m$ and the same total harvesting ability $\sum_{\alpha=1}^M c_{i\alpha} = c$ (\citealt{macarthur1970species}). This same constraint has been discussed recently in the context of microbial ecology (\citealt{posfai2017metabolic}), where it has been shown to give rise to non-generic behavior in highly diverse communities (\citealt{cui2019effect}). In this scenario, MacArthur's objective function from eq.~(\ref{eq:QN}) in the main text can be written as:
\begin{align}
Q(\mathbf{N}) &= \frac{1}{2} \sum_{\alpha \in \mathbf{M}^*}  \frac{K_\alpha}{r_\alpha} w_\alpha \left[\frac{r_\alpha}{K_\alpha}\left(K_\alpha -  \frac{m}{c w_\alpha}\right)-\sum_j c_{j\alpha} N_j\right]^2 + \frac{m}{c}\left(\sum_{\alpha \notin \mathbf{M}^*} c_{j\alpha}\right)\sum_{j} N_j
\end{align}
where $\mathbf{M}^*$ is the set of resources where $r_\alpha \geq \sum_j c_{j\alpha} N_j$, which can stably avoid extinction at the current consumer population size. 

When all the resource types survive, the final term in this expression vanishes, and the remaining part takes on the straightforward physical meaning proposed by MacArthur. $\frac{r_\alpha}{K_\alpha}\left(K_\alpha -  \frac{m}{c w_\alpha}\right)$ is the production rate of resource $\alpha$ when the abundance $R_\alpha$ is at the minimum value that supports consumer growth. The objective function is a weighted sum of squared differences between this ``available production'' and the community's total harvest rate $\sum_j c_{j\alpha} N_j$. These ecological dynamics can thus be conceived of as an algorithm for performing a least-squares fit of the harvest rate (with positive free parameters $N_j$) to the available production. 

But if any resources go extinct in the steady state, this interpretation is no longer valid. Now some terms end up disappearing from the first sum, with corresponding modifications to the final term, which has no clear biological meaning. Even in this case, however, a revised explanation by M. Gatto still applies (\citealt{gatto1990general}). In this reading, no constraints on $m_i$ or $c_{i\alpha}$ are required, and one instead directly interprets the two terms that already appeared in the original expression for $Q$ in eq.~(\ref{eq:QN}) The first term, which he calls the ``unutilized productivity'' $U$, is a weighted sum of squared differences between the maximal resource production rate $r_\alpha$ and the current consumption rate:
\begin{align}
U =  \frac{1}{2} \sum_{\alpha \in \mathbf{M}^*} r_\alpha^{-1} K_\alpha w_\alpha \left(r_\alpha - \sum_j c_{j\alpha} N_j \right)^2.
\end{align}
While Gatto does not comment on the restriction of the sum to the surviving resources, this interpretation of $U$ is compatible with the restriction. If a resource is extinct, it is reasonable to say that none of its (nonexistent) potential productivity is unutilized. The second term is the ``basal energy consumption'' $B$ which is the total consumption of nutritional value by the community required to maintain the current population sizes:
\begin{align}
B = \sum_j m_j N_j.
\end{align}
This term is not affected by resource extinction, and the interpretation remains valid. 

The full expression for $Q$ in eq.~(\ref{eq:QN}) can also be rearranged in a different way, which sets the stage for the present work. To obtain this form, we first note that the local equilibrium abundance $\bar{R}_\alpha$ of resource $\alpha$ at fixed consumer population sizes $N_i$ are given by
\begin{align}
\bar{R}_\alpha(\mathbf{N}) = {\rm max}\left[0, K_\alpha \left(1-r_\alpha^{-1} \sum_i N_i c_{i\alpha}\right)\right].
\end{align}
This expression comes from the fact that there are two solutions to $dR_\alpha/dt =0$, one where $R_\alpha = 0$ and one given by the second term in the brackets. Since resource abundances must be positive, we are required to take $R_\alpha = 0$ if the nonzero solution turns out to be negative. If the nonzero solution is positive, then the $R_\alpha = 0$ solution is unstable to the addition of a small amount of resource $\alpha$. This consideration fully accounts for resource extinction, and so eq.~(\ref{eq:QN}) simplifies to
\begin{align}
Q(\mathbf{N}) = -\frac{1}{2} \sum_{\alpha} \frac{w_\alpha r_\alpha}{K_\alpha} [K_\alpha - \bar{R}_\alpha(\mathbf{N})]^2 - \sum_i N_i \left(\sum_\alpha w_\alpha c_{i\alpha} \bar{R}_\alpha(\mathbf{N}) - m_i \right)
\end{align}
where the sums are no longer restricted. The first term now measures the difference between the local equilibrium resource concentrations $\bar{R}_\alpha(\mathbf{N})$ and the carrying capacities $K_\alpha$, while the second term measures the total rate of biomass production. This form of $Q(\mathbf{N})$ also makes it easier to see that MacArthur's minimization principle is the Lagrange dual of MEPP (\citealt{boyd2004convex}). The first term is clearly minus the objective function $d$ defined in eq.~(\ref{eq:dmcrm}), and the second term is the sum of the Lagrange multipliers times the active constraints $g_i$, with $R_\alpha$ replaced by $\bar{R}_\alpha(\mathbf{N})$ in both terms. 

\section*{Appendix B: Derivation of Minimum Environmental Perturbation Principle}
In this Appendix, we justify the three mathematical results required for the derivation of MEPP in the main text: 
\begin{itemize}
	\item that the impact vectors are related to the gradients of the growth rates by eq.~(\ref{eq:cond1}) whenever the environmentally mediated interactions between species are symmetric
	\item that this same symmetry implies that the rescaled supply vector $h_\alpha/b_\alpha$ can be written as the (negative) gradient of some function $d$, as done in eq.~(\ref{eq:dddR})
	\item that the unconstrained minimum of $d$ coincides with the supply point of the resource dynamics.
\end{itemize}
We also explain how an extended version of MEPP can be obtained for asymmetric models by using a modified supply vector.

\subsection*{Implications of symmetric interactions}
In this section we deal with the first two points in the list, concerning the consequences of symmetric interactions. To quantify the interactions between two species, we compute the effect of a small change in the abundance of the first species on the growth rate of the second. We introduce a scale factor $a_i$ that can depend on the environmental state, and measure abundances as $a_i N_i$. Since the growth rates directly depend only on the resource abundances, we need to imagine making the perturbation and then holding all the population sizes fixed until the environment relaxes to its new equilibrium state $\bar{\mathbf{R}}(\mathbf{N})$. Thus we define the interaction matrix $\alpha_{ij}$ as:
\begin{align}
\alpha_{ij}&= -\frac{d g_i}{d (a_j N_j)} = -\sum_\alpha \frac{\partial g_i}{\partial R_\alpha}\frac{\partial \bar{R}_\alpha}{\partial (a_jN_j)}.
\end{align}
Now we can compute $\frac{\partial \bar{R}_\alpha}{\partial (a_jN_j)}$ by implicit differentiation of the local steady-state equation for the environment, and thus obtain an explicit expression for $\alpha_{ij}$. Setting $dR_\alpha/dt = 0$ in eq.~(\ref{eq:niche2}) yields:
\begin{align}
0 &= h_\alpha + \sum_i N_i q_{i\alpha}\\
&= h_\alpha - \sum_i N_i a_i \sum_\beta b^i_{\alpha\beta} \frac{\partial g_i}{\partial R_\beta}.
\end{align}
In the second line we have written $q_{i\alpha}$ as $-a_i \sum_\beta b^i_{\alpha\beta} \frac{\partial g_i}{\partial R_\beta}$. This is only a notational convenience for subsequent steps of the derivation, but does not impose any additional assumptions on the form of $q_{i\alpha}$, as long as $\frac{\partial g_i}{\partial R_\beta}\neq 0$. If we now further assume that the $b^i_{\alpha\beta}$ are invertible, we can multiply by $(b^j)^{-1}$ and obtain:
\begin{align}
0 &= \sum_{\beta} (b^j)^{-1}_{\alpha\beta} h_\beta - \sum_{i\beta\gamma} N_i a_i (b^j)^{-1}_{\alpha\beta} b^i_{\beta\gamma}\frac{\partial g_i}{\partial R_\gamma}.
\end{align}
Taking the derivative of both sides with respect to $a_j N_j$ gives:
\begin{align}
0 &= \sum_{\lambda} \frac{\partial}{R_\lambda}\left(\sum_\beta (b^j)^{-1}_{\alpha\beta} h_\beta\right)\frac{\partial \bar{R}_\lambda}{\partial (a_j N_j)}  - \sum_{i\beta\gamma\lambda} N_i a_i (b^j)^{-1}_{\alpha\beta} b^i_{\beta\gamma} \frac{\partial^2 g_i}{\partial R_\lambda \partial R_\alpha}\frac{\partial \bar{R}_\lambda}{\partial (a_j N_j)} - \frac{\partial g_j}{\partial R_\alpha}\\
&= -\sum_{\lambda} A_{\alpha\lambda}^j \frac{\partial \bar{R}_\lambda}{\partial (a_j N_j)} - \frac{\partial g_j}{\partial R_\alpha}
\end{align}
where
\begin{align}
A_{\alpha\lambda}^j =-\frac{\partial}{\partial R_\lambda}\left(\sum_\beta (b^j)^{-1}_{\alpha\beta} h_\beta\right)  + \sum_{i\beta\gamma} N_i a_i (b^j)^{-1}_{\alpha\beta} b^i_{\beta\gamma}\frac{\partial^2 g_i}{\partial R_\lambda \partial R_\alpha}.
\label{eq:A1}
\end{align}
Now, further assuming that this matrix is invertible, we obtain:
\begin{align}
\frac{\partial \bar{R}_\alpha}{\partial N_j} = -\sum_\beta (A^j)^{-1}_{\alpha \beta}\frac{\partial g_j}{\partial R_\beta}.
\end{align}
Finally, inserting this into the definition of the interaction matrix yields
\begin{align}
\alpha_{ij}&= \sum_{\alpha\beta} (A^j)^{-1}_{\alpha\beta} \frac{\partial g_i}{\partial R_\alpha}\frac{\partial g_j}{\partial R_\beta}.
\end{align}

With this expression in hand, we can proceed to investigate the implications of symmetry ($\alpha_{ij} = \alpha_{ji}$), by looking for conditions under which
\begin{align}
\sum_{\alpha\beta} (A^j)^{-1}_{\alpha\beta} \frac{\partial g_i}{\partial R_\alpha}\frac{\partial g_j}{\partial R_\beta} = \sum_{\alpha\beta} (A^i)^{-1}_{\alpha\beta} \frac{\partial g_j}{\partial R_\alpha}\frac{\partial g_i}{\partial R_\beta}.
\end{align}
Inspection of this equation reveals two important conditions. The first is that $A^j_{\alpha\beta}$ is the same for all $j$. Going back to the definition of $A^j_{\alpha\beta}$ in eq.~(\ref{eq:A1}), we find that this is true if and only if $b^j_{\alpha\beta}$ is the same for all $j$. In this case, the definition simplifies to
\begin{align}
A_{\alpha\lambda} =-\frac{\partial}{\partial R_\lambda}\left(\sum_\beta b^{-1}_{\alpha\beta} h_\beta\right)  + \sum_{i} N_i a_i \frac{\partial^2 g_i}{\partial R_\lambda \partial R_\alpha}.
\label{eq:A2}
\end{align}
The second condition is that $A_{\alpha\beta}$ must itself be symmetric. The second term in eq.~(\ref{eq:A2}) is always symmetric, so we can focus on the first. Symmetry of this term means that
\begin{align}
\frac{\partial}{\partial R_\lambda}\left(\sum_\beta b^{-1}_{\alpha\beta} h_\beta\right) = \frac{\partial}{\partial R_\alpha}\left(\sum_\beta b^{-1}_{\lambda\beta} h_\beta\right).
\end{align}
For this to be satisfied in a generic model, $b_{\alpha\beta}$ must be diagonal, so that the requirement becomes:
\begin{align}
\frac{\partial}{\partial R_\lambda}\frac{h_\alpha}{b_\alpha}= \frac{\partial}{\partial R_\alpha}\frac{h_\lambda}{b_\lambda},
\end{align}
using $b_{\alpha\beta} = b_\alpha \delta_{\alpha\beta}$. If $b_{\alpha\beta}$ is not diagonal, very specific correlations between the $\mathbf{R}$-dependence of $b^{-1}_{\alpha\beta}$ and the $h_\alpha$ would be required to satisfy the condition. This simplified version is sufficient to guarantee that $\frac{h_\alpha}{b_\alpha}$ can be written as a gradient of some function, as claimed in the main text:
\begin{align}
\frac{\partial d}{\partial R_\alpha} &= -\frac{h_\alpha}{b_\alpha}.
\end{align}
In this case, $A_{\alpha\beta}$ further simplifies to:
\begin{align}
A_{\alpha\lambda} = \frac{\partial^2 d}{\partial R_\lambda\partial R_\alpha} + \sum_{i} N_i a_i \frac{\partial^2 g_i}{\partial R_\lambda \partial R_\alpha}.
\label{eq:A3}
\end{align}

We can restate these corollaries of interaction symmetry in a particularly useful way by returning to the dynamical equations. The preceding arguments show that the environmentally mediated interactions between species in a generic niche model described by eqs.~(\ref{eq:niche1}-\ref{eq:niche2}) are symmetric if and only if the dynamics can be rewritten as:
\begin{align}
\frac{dN_i}{dt} &= N_i g_i(\mathbf{R})\\
\frac{dR_\alpha}{dt} &= -b_\alpha \left[ \frac{\partial d}{\partial R_\alpha} + \sum_i a_i N_i \frac{\partial g_i}{\partial R_\alpha}\right]\label{eq:r2}
\end{align}
for some functions $b_{\alpha}(\mathbf{R})$ and $a_i(\mathbf{R})$.

\subsection*{Supply point as unconstrained minimum}
In the main text, we made the assumption that $b_\alpha> 0$, and that the supply point $\mathbf{R}^0$ is a stable fixed point of the intrinsic environmental dynamics $\frac{dR_\alpha}{dt} = h_\alpha(\mathbf{R})$. We evaluate the stability of the fixed point in the usual way, by computing the Jacobian $\frac{\partial h_\alpha}{\partial R_\beta}$. The equilibrium point is stable if and only if this matrix is negative definite, so that the dynamics tend to resist small perturbations from equilibrium. Now from the definition of $d$ in eq.~(\ref{eq:dddR}) we have
\begin{align}
\frac{\partial h_\alpha}{\partial R_\beta} = - \frac{\partial^2 d}{\partial R_\beta\partial R_\alpha} b_\alpha - \frac{\partial d}{\partial R_\alpha}\frac{\partial b_\alpha}{\partial R_\beta}
\end{align}
where the second term vanishes at the supply point $\mathbf{R}^0$ since $h_\alpha = -b_\alpha \frac{\partial d}{\partial R_\alpha} = 0$ there. From the remaining term and the fact that $b_\alpha >0$, standard results on D-stability (cf. \citealt{hogben2013handbook}) yield that the Hessian $\frac{\partial^2 d}{\partial R_\beta\partial R_\alpha}$ is positive definite whenever $\partial h_\alpha/\partial R_\beta$ is negative definite. Thus we arrive at the result stated in the main text, that the supply point $\mathbf{R}^0$ is an unconstrained local minimum of $d$.

\subsection*{Extended MEPP for arbitrary niche models}
Here we show how to obtain and use a minimization principle for models with asymmetric interactions between species, where the impact vector and growth rate cannot be related by an equation of the form of eq.~(\ref{eq:dddR}). We do this by constructing a symmetric model that shares the same stable equilibrium point $\bar{\mathbf{N}},\bar{\mathbf{R}}$. The equilibrium condition of the original dynamics is:
\begin{align}
0 = h_\alpha(\bar{\mathbf{R}}) + \sum_i \bar{N}_i q_{i\alpha}(\bar{\mathbf{R}}).
\end{align}
Now for any positive functions $b_\alpha(\mathbf{R})$ and $a_i(\mathbf{R})$, and any convex function $d(\mathbf{R})$ we can trivially write
\begin{align}
0 = -b_\alpha\left[ \frac{\partial d}{\partial R_\alpha} - \sum_i a_i \bar{N}_i \frac{\partial g_i}{\partial R_\alpha}\right] +  \left[h_\alpha + \sum_i \bar{N}_i q_{i\alpha} +b_\alpha \frac{\partial d}{\partial R_\alpha}+ \sum_i a_i \bar{N}_i b_\alpha \frac{\partial g_i}{\partial R_\alpha}\right] \label{eq:dyncon}
\end{align}
where all functions are evaluated at $\bar{\mathbf{R}}$. Finally, we choose $d$ in such a way that the quantity in the second set of brackets vanishes at the equilibrium point $\bar{\mathbf{N}},\bar{\mathbf{R}}$:
\begin{align}
\frac{\partial d}{\partial R_\alpha} &= -\frac{1}{b_\alpha}\left [h_\alpha + \sum_i \bar{N}_i q_{i\alpha} +\sum_i a_i \bar{N}_i b_\alpha \frac{\partial g_i}{\partial R_\alpha}\right]\label{eq:dmod}
\end{align}
This is only possible if we have that ${\partial \over \partial R_\beta} \frac{\partial d}{\partial R_\alpha}= {\partial \over \partial R_\alpha} \frac{\partial d}{\partial R_\beta}$.Since $\bar{\mathbf{N}}$ and $\bar{\mathbf{R}}$ are independent of $R_\alpha$ one
can verify the this equation is satisfied by the distance function 
 \begin{align}
d=-\sum_\alpha R_\alpha h_\alpha(\bar{\mathbf{R}})/b_\alpha(\bar{\mathbf{R}})+\sum_i a_i \bar{N}_i b_\alpha g_i(\mathbf{R})
\end{align}
With this choice of $d$, eq.~(\ref{eq:dyncon}) is also the equilibrium condition for the symmetric model with environmental dynamics given by 

\begin{align}
\frac{dR_\alpha}{dt} = -b_\alpha\left[ \frac{\partial d}{\partial R_\alpha} - \sum_i a_i N_i \frac{\partial g_i}{\partial R_\alpha}\right]
\end{align}
which is guaranteed to minimize $d$ subject to the constraints $g_i \leq 0$ for all species $i$. 

One important special case of this general procedure is when the asymmetric model can be constructed from a reference symmetric model by an additive modification to the impact vector. In this case, we can write the impact vector as
\begin{align}
q_{i\alpha} = q_{i\alpha}^S + q_{i\alpha}^A
\end{align}
where $q_{i\alpha}^S$ is the impact vector from the symmetric model and $q_{i\alpha}^A$ is the modification. Substituting in to the general equation for the resource dynamics (\ref{eq:niche2}) and rearranging, we obtain
\begin{align}
\frac{dR_\alpha}{dt} &= h_\alpha - \sum_i N_i q_{i\alpha}^A(\mathbf{R}) + \sum_i N_i q_{i\alpha}^S(\mathbf{R}).
\end{align}
We can now obtain a symmetric model that shares the same equilibrium state $(\bar{\mathbf{N}},\bar{\mathbf{R}})$ as the original model by simply replacing $N_i$ and $\mathbf{R}$ with their equilibrium values $\bar{N}_i$ and $\bar{\mathbf{R}}$ in the sum over the asymmetric parts of the impact vectors, so that
\begin{align}
\frac{dR_\alpha}{dt} &= \tilde{h}_\alpha + \sum_i N_i q_{i\alpha}^S(\mathbf{R}).
\end{align}
with supply vector
\begin{align}
\tilde{h}_\alpha = h_\alpha - \sum_i \bar{N}_i q_{i\alpha}^A(\bar{\mathbf{R}}).\label{eq:htild}
\end{align}
We can now write down the expression for the objective function using the general formula for symmetric models:
\begin{align}
\frac{\partial d}{\partial R_\alpha} = - \frac{\tilde{h}_\alpha}{b_\alpha}.\label{eq:dmod2}
\end{align}

As noted in the main text, the problem with the extended version of MEPP is that the construction of $d$ requires prior knowledge of the equilibrium state $\bar{\mathbf{N}}, \bar{\mathbf{R}}$. This strange problem of minimizing an objective function whose parameters depend on the solution arises frequently in Machine Learning, in the context of fitting models with latent variables (\citealt{mehta2018high}). It can be solved with a simple iterative approach, called Expectation Maximization, where one starts by guessing the values of these parameters, then minimizes the function, and then updates the estimates using the new solution:
\begin{enumerate}
	\item Initialize $\bar{\mathbf{N}}, \bar{\mathbf{R}}$ with arbitrarily chosen values
	\item Compute $d$ using eq.~(\ref{eq:dmod}) and the current estimate of $\bar{\mathbf{N}}, \bar{\mathbf{R}}$.
	\item Minimize $d$ to update estimate of $\bar{\mathbf{N}}, \bar{\mathbf{R}}$.
	\item Repeat steps 2-3 until the estimate of $\bar{\mathbf{N}},\bar{\mathbf{R}}$ stops changing. 
\end{enumerate}
It is possible for this algorithm to fail, if the estimate $\bar{\mathbf{N}},\bar{\mathbf{R}}$ never stops changing. But if the algorithm does converge, it clearly solves the correct optimization problem, minimizing $d$ using the true value of $\bar{\mathbf{R}}$. Fig. \ref{fig:asymmetric} compares the output of this algorithm with direct numerical simulation of the two asymmetric examples.

\section*{Appendix C: Analysis of specific models} 
In this Appendix, we show in detail how to obtain the objective function $d$ and the auxiliary functions $b_\alpha$ and $a_i$ for each of the seven models analyzed here. We do this by following the first two steps of the procedure outlined in the main text, which are copied here for reference:
\begin{enumerate}
	\item Find $b_\alpha$ and $a_i$ by comparing the impact vectors with the derivative of the growth rates using $q_{i\alpha}(\mathbf{R}) = - a_i(\mathbf{R}) b_{\alpha}(\mathbf{R}) \frac{\partial g_i}{\partial R_\alpha}$.
	\item Compute $d$ from $b_\alpha$ and the supply vector using $\frac{\partial d}{\partial R_\alpha} = -\frac{h_\alpha(\mathbf{R})}{b_\alpha(\mathbf{R})}$.
\end{enumerate}

\begin{figure*}
	\includegraphics[width=16cm]{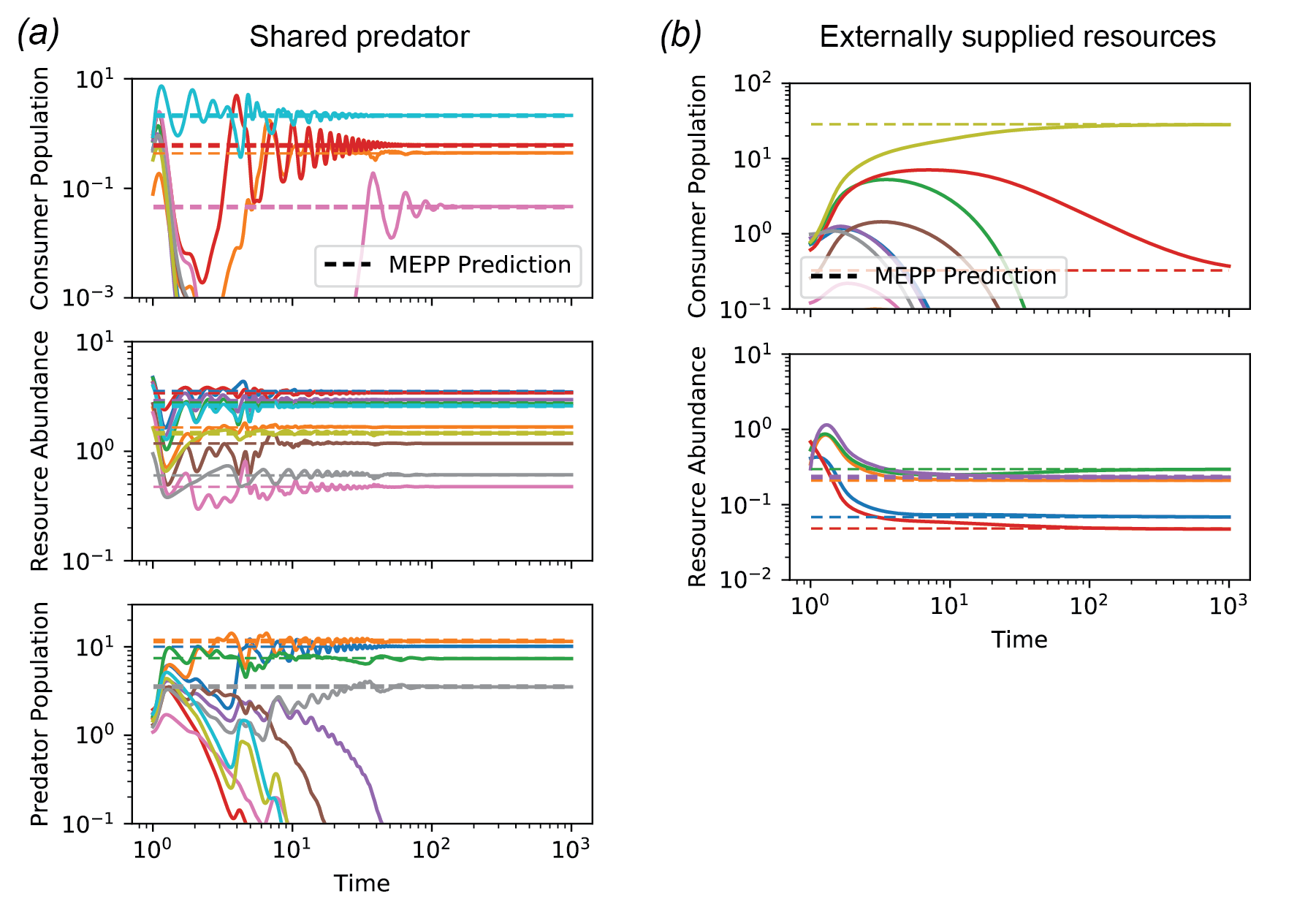}
	\caption{\linespread{1.3}\selectfont{} {\bf Additional simulations with symmetric interactions.} Simulations of the second two models from fig.~\ref{fig:symmetric} with larger numbers of species and resources, compared with the predictions of MEPP for the uninvadable equilibrium state. Consumer abundances are obtained from the Lagrange multipliers that enforce the constraints during optimization. See Appendix text or Jupyter notebooks for all simulation parameters. }
	\label{fig:extra}
\end{figure*}

\subsection*{Noninteracting resources}
We begin with the dynamical equations
\begin{align}
\frac{dN_i}{dt} &= e_i N_i \left[ \sum_{\alpha} w_\alpha c_{i\alpha} R_\alpha - m_i\right]\\
\frac{dR_\alpha}{dt} &= \frac{r_\alpha}{K_\alpha} R_\alpha (K_\alpha - R_\alpha) - \sum_i N_i c_{i\alpha} R_\alpha.\label{eq:MacR2}
\end{align}
Comparing with the general niche theory scheme of eq.~(\ref{eq:niche1}-\ref{eq:niche2}), we identify
\begin{align}
g_i(\mathbf{R}) &= e_i \left[ \sum_{\alpha} w_\alpha c_{i\alpha} R_\alpha - m_i\right]\\
q_{i\alpha}(\mathbf{R}) &= - c_{i\alpha} R_\alpha\\
h_\alpha(\mathbf{R}) &= \frac{r_\alpha}{K_\alpha} R_\alpha (K_\alpha - R_\alpha),
\end{align}
as also given in table \ref{tab:niche}. The gradient of the growth rate is
\begin{align}
\frac{\partial g_i}{\partial R_\alpha} = e_i w_\alpha c_{i\alpha}.
\end{align}
Now we can follow Step 1 from the list above, comparing this to the impact vector to obtain
\begin{align}
a_i &= e_i^{-1}\\
b_\alpha &= \frac{R_\alpha}{w_\alpha}.
\end{align}
Step 2 now yields the expression for $d$:
\begin{align}
\frac{\partial d}{\partial R_\alpha} = - \frac{r_\alpha w_\alpha}{K_\alpha} (K_\alpha - R_\alpha).
\end{align}
Integrating this expression, we find
\begin{align}
d = \frac{1}{2}\sum_\alpha \frac{r_\alpha w_\alpha}{K_\alpha}(K_\alpha - R_\alpha)^2
\end{align}
which is equivalent to eq.~(\ref{eq:dmcrm}) in the main text.

We are not quite finished, however, because the minimization of this expression for $d$ subject to $g_i \leq 0$ can produce negative values of $\bar{R}_\alpha$. Physically, we know that the resource abundances cannot be negative numbers, and the original dynamical equation (\ref{eq:MacR2}) ensures that $R_\alpha$ never becomes negative as long as the initial conditions are positive. But this constraint is lost when we divide by $b_\alpha = R_\alpha/w_\alpha$ in the derivation of the KKT conditions. To address this issue, one must impose $R_\alpha \geq 0$ as an additional set of explicit constraints when performing the optimization. This problem occurs for most models with self-renewing resources, and can always be resolved by adding additional constraints in this way.

\subsection*{Interacting self-regulation}
We begin with the dynamical equations
\begin{align}
\frac{dN_i}{dt} &= e_i N_i \left[ \sum_{\alpha} w_\alpha c_{i\alpha} R_\alpha - m_i\right]\\
\frac{dR_\alpha}{dt} &= r_\alpha R_\alpha \left( 1- \sum_\beta a_\beta R_\beta\right) - \sum_i N_i c_{i\alpha} R_\alpha.
\end{align}
Comparing with the general niche theory scheme of eq.~(\ref{eq:niche1}-\ref{eq:niche2}), we make the same identifications as for noninteracting resources, but with a modified supply vector:
\begin{align}
g_i(\mathbf{R}) &= e_i \left[ \sum_{\alpha} w_\alpha c_{i\alpha} R_\alpha - m_i\right]\\
q_{i\alpha}(\mathbf{R}) &= - c_{i\alpha} R_\alpha\\
h_\alpha(\mathbf{R}) &=r_\alpha R_\alpha \left( 1- \sum_\beta a_\beta R_\beta\right).
\end{align}
Since $g_i$ and $q_{i\alpha}$ are unchanged, we have the same expressions for $a_i$ and $b_\alpha$:
\begin{align}
a_i &= e_i^{-1}\\
b_\alpha &= \frac{R_\alpha}{w_\alpha}.
\end{align}
Step 2 now yields the expression for $d$:
\begin{align}
\frac{\partial d}{\partial R_\alpha} = - r_\alpha w_\alpha \left( 1- \sum_\beta a_\beta R_\beta\right).
\end{align}
For generic $a_\beta$ and $w_\alpha$, there is no function $d$ that satisfies this expression, because the second derivatives of the function would be:
\begin{align}
\frac{\partial^2 d}{\partial R_\beta \partial R_\alpha} = - r_\alpha w_\alpha a_\beta \neq \frac{\partial^2 d}{\partial R_\alpha \partial R_\beta}.
\end{align}
This means that the model is generically not symmetric. But if we set $w_\alpha = w a_\alpha$ as described in the main text, we find that 
\begin{align}
d = \frac{wr}{2}\left( 1- \sum_\alpha a_\alpha R_\alpha\right)^2
\end{align}
satisfies the equation. 

\subsection*{Shared predators}
We begin with the dynamical equations
\begin{align}
\frac{dN_i}{dt} &= e_i N_i \left[ \sum_{\alpha} w_\alpha c_{i\alpha} R_\alpha - m_i \right] - \sum_a p_{ia} P_a N_i\\
\frac{dR_\alpha}{dt} &= \frac{r_\alpha}{K_\alpha} R_\alpha (K_\alpha - R_\alpha) - \sum_i N_i c_{i\alpha} R_\alpha\\
\frac{dP_a}{dt} &= \sum_i \eta_i p_{ia} N_i P_a - u_a P_a
\end{align}
To situate this model within the general niche theory scheme of eq.~(\ref{eq:niche1}-\ref{eq:niche2}), we must treat the predators as additional environmental factors, along with the resources. We denote the impact and supply vectors for the resources by $q_{i\alpha}^R$ and $h_\alpha^R$, and the corresponding vectors for the predators by $q_{ia}^P$ and $h_a^p$. We obtain:
\begin{align}
g_i(\mathbf{R},\mathbf{P}) &= e_i \left[ \sum_{\alpha} w_\alpha c_{i\alpha} R_\alpha - m_i \right] - \sum_a p_{ia} P_a\\
q_{i\alpha}^R(\mathbf{R}) &= - c_{i\alpha} R_\alpha\\
q_{ia}^P(\mathbf{P}) &= \eta_i p_{ia} P_a\\
h_\alpha^R(\mathbf{R}) &=\frac{r_\alpha}{K_\alpha} R_\alpha (K_\alpha - R_\alpha)\\
h_a^P(\mathbf{P}) &=- u_a P_a.
\end{align}
Following Step 1 from the general procedure with $q_{i\alpha}^R$ as the impact vector yields the same results for $a_i$ and $b_\alpha$ as the previous two cases, while using $q_{ia}^P$ yields:
\begin{align}
a_i &= \frac{\eta_i}{e_0}\\
b_a^P &= e_0 P_a
\end{align}
for an arbitrary constant $e_0$, which we will have need of soon. We have added a superscript to $b_a^P$, because there is a separate set of these functions for the predators and for the resources. In the formula for the impact vector in Step 1, however, there can only be one value of $a_i$ per consumer species. This means that this formula can only be satisfied if
\begin{align}
e_i = \frac{e_0}{\eta_i}
\end{align}
which is the requirement for symmetric interactions stated in the main text.

Under this assumption, we can apply Step 2 to obtain expressions for the derivatives of $d$:
\begin{align}
\frac{\partial d}{\partial R_\alpha} &= - \frac{r_\alpha}{K_\alpha} w_\alpha(K_\alpha - R_\alpha)\\
\frac{\partial d}{\partial P_a} &= \frac{u_a}{e_0}.
\end{align}
Integrating these expressions, we obtain:
\begin{align}
d &= \frac{1}{2}\sum_\alpha \frac{r_\alpha}{K_\alpha} w_\alpha (K_\alpha - R_\alpha)^2 + \frac{1}{e_0} \sum_a u_a P_a
\end{align}
as reported in the main text.

\subsection*{Externally supplied resources}
We begin with the dynamical equations
\begin{align}
\frac{dN_i}{dt} &= e_i N_i \left[ \sum_{\alpha} w_\alpha c_{i\alpha} R_\alpha - m_i\right]-\tau^{-1} N_i\label{eq:extN2}\\
\frac{dR_\alpha}{dt} &= \tau^{-1} (R_\alpha^0 - R_\alpha) - \sum_i N_i c_{i\alpha} R_\alpha.\label{eq:extR2}
\end{align}
Comparing with the general niche theory scheme of eq.~(\ref{eq:niche1}-\ref{eq:niche2}), we identify
\begin{align}
g_i(\mathbf{R}) &= e_i \left[ \sum_{\alpha} w_\alpha c_{i\alpha} R_\alpha - m_i\right]-\tau^{-1}\\
q_{i\alpha}(\mathbf{R}) &= - c_{i\alpha} R_\alpha\\
h_\alpha(\mathbf{R}) &= \tau^{-1}(R_\alpha^0-R_\alpha),
\end{align}
which is the same as for the original consumer resource model (\ref{eq:MacN}-\ref{eq:MacR}), except for the supply vector. We thus obtain the same conversion factors:
\begin{align}
a_i &= e_i^{-1}\\
b_\alpha &= \frac{R_\alpha}{w_\alpha}.
\end{align}
Step 2 now yields the expression for $d$:
\begin{align}
\frac{\partial d}{\partial R_\alpha} = -\tau^{-1}w_\alpha \frac{R_\alpha^0 - R_\alpha}{R_\alpha}
\end{align}
Integrating this expression, we find
\begin{align}
d = \tau^{-1}  \sum_\alpha w_\alpha\left[R_\alpha^0 \ln \frac{R_\alpha^0}{R_\alpha} - (R_\alpha^0 - R_\alpha)\right]. 
\end{align}
which is eq.~(\ref{eq:KL}) in the main text. Note that this expression diverges as $R_\alpha \to 0$, so there is no need to explicitly impose the resource feasibility constraints.

\subsection*{Microbial consumer resource model}
We begin with the dynamical equations
\begin{align}
\frac{dN_i}{dt} &= e_i N_i \left[ \sum_{\alpha} (1-l_\alpha)w_\alpha c_{i\alpha} R_\alpha - m_i\right]-\tau^{-1} N_i\\
\frac{dR_\alpha}{dt} &= \tau^{-1} (R_\alpha^0 - R_\alpha) - \sum_i N_i c_{i\alpha} R_\alpha + \sum_{i\beta} N_i D_{\alpha\beta}l_\beta \frac{w_\beta}{w_\alpha}c_{i\beta}R_\beta.\label{eq:micR2}
\end{align}
Comparing with the general niche theory scheme of eq.~(\ref{eq:niche1}-\ref{eq:niche2}), we identify
\begin{align}
g_i(\mathbf{R}) &= e_i \left[ \sum_{\alpha} (1-l_\alpha) w_\alpha c_{i\alpha} R_\alpha - m_i\right]-\tau^{-1}\\
q_{i\alpha}(\mathbf{R}) &= - c_{i\alpha} R_\alpha +\sum_{\beta} D_{\alpha\beta}l_\beta \frac{w_\beta}{w_\alpha}c_{i\beta}R_\beta\\
h_\alpha(\mathbf{R}) &= \tau^{-1}(R_\alpha^0-R_\alpha).
\end{align}
As noted in the main text, the generation of byproducts breaks the symmetry of interactions between consumers, and so we must use the extended form of MEPP discussed above. Since all the asymmetry comes from the production part of the impact vector, we can follow the simplified procedure based on a splitting of the impact vector $q_{i\alpha} = q_{i\alpha}^S + q_{i\alpha}^A$. In this case, the symmetric reference model has an impact vector $q_{i\alpha}^S$ identical to that of an ordinary consumer-resource model, and $q_{i\alpha}^A$ encodes byproduct generation:
\begin{align}
q_{i\alpha}^S &= -c_{i\alpha}R_\alpha\\
q_{i\alpha}^A &= \sum_{\beta} D_{\alpha\beta}l_\beta \frac{w_\beta}{w_\alpha}c_{i\beta}R_\beta.
\end{align} 
We thus see that this model shares an equilibrium state with a pure competition model of the form (\ref{eq:extN2}-\ref{eq:extR2}), but with a modified supply vector
\begin{align}
\tilde{h}_\alpha &= \tau^{-1}(R_\alpha^0-R_\alpha) + \sum_{i\beta} \bar{N}_i D_{\alpha\beta}l_\beta \frac{w_\beta}{w_\alpha}c_{i\beta}\bar{R}_\beta
\end{align}
and modified resource weights
\begin{align}
\tilde{w}_\alpha = w_\alpha (1-l_\alpha).
\end{align}
The change to the supply vector is equivalent to a shift of the supply point from $R_\alpha^0$ to
\begin{align}
\tilde{R}_\alpha^0 = R_\alpha^0 + \tau \sum_{i\beta} \bar{N}_i D_{\alpha\beta}l_\beta \frac{w_\beta}{w_\alpha}c_{i\beta}\bar{R}_\beta,
\end{align}
which accounts for the total quantity of byproducts generated by all consumers over one chemostat turnover time $\tau$.

We can therefore use the same objective function obtained for the pure competition model in eq.~(\ref{eq:KL}), but with these modified formulas for the weights $w_\alpha$ and the supply point $\tilde{R}_\alpha^0$.

\subsection*{Liebig's Law}
We begin with the dynamical equations
\begin{align}
\frac{dN_i}{dt} &= N_i \left[\underset{\beta}{\rm min}\left(\left\{\frac{\mu_{i\beta} R_\beta}{k_{i\beta}+R_\beta}\right\}\right) - m_i \right]  \label{eq:lie12}\\
\frac{dR_\alpha}{dt} &= \tau^{-1} (R_\alpha^0 - R_\alpha) - \sum_{i} N_i \nu_{\alpha i} \underset{\beta}{\rm min}\left(\left\{\frac{\mu_{i\beta} R_\beta}{k_{i\beta}+R_\beta}\right\}\right)\label{eq:lie22}.
\end{align}
Comparing with the general niche theory scheme of eq.~(\ref{eq:niche1}-\ref{eq:niche2}), we identify
\begin{align}
g_i(\mathbf{R}) &=\underset{\beta}{\rm min}\left(\left\{\frac{\mu_{i\beta} R_\beta}{k_{i\beta}+R_\beta}\right\}\right) - m_i \\
q_{i\alpha}(\mathbf{R}) &= - \nu_{\alpha i} \underset{\beta}{\rm min}\left(\left\{\frac{\mu_{i\beta} R_\beta}{k_{i\beta}+R_\beta}\right\}\right)\\
h_\alpha(\mathbf{R}) &= \tau^{-1}(R_\alpha^0-R_\alpha),
\end{align}
As noted in the main text, the consumption of resources that are not currently limiting growth breaks the symmetry of the interactions between consumers, and so we must use the extended form of MEPP discussed above. Since all the asymmetry comes from this ``excess'' consumption, we can follow the simplified procedure based on a splitting of the impact vector $q_{i\alpha} = q_{i\alpha}^S + q_{i\alpha}^A$. In this case, the symmetric reference model has an impact vector $q_{i\alpha}^S$ that only depletes the limiting nutrient, and $q_{i\alpha}^A$ encodes the consumption of non-limiting nutrients. To write explicit expressions for these quantities, it is convenient to denote the index of the limiting resource by $\beta_i$, so that $\underset{\beta}{\rm min}\left(\left\{\frac{\mu_{i\beta} R_\beta}{k_{i\beta}+R_\beta}\right\}\right) = \frac{\mu_{i\beta_i} R_{\beta_i}}{k_{i\beta_i}+R_{\beta_i}}$. Then we have
\begin{align}
q_{i\alpha}^S &= - \nu_{\beta_i i}\delta_{\alpha\beta_i} \frac{\mu_{i\beta_i} R_{\beta_i}}{k_{i\beta_i}+R_{\beta_i}}\\
q_{i\alpha}^A &=- \nu_{\alpha i} (1-\delta_{\alpha\beta_i}) \frac{\mu_{i\beta_i} R_{\beta_i}}{k_{i\beta_i}+R_{\beta_i}}.
\end{align} 
We obtain $a_i$ and $b_\alpha$ by comparing $q_{i\alpha}^S$ with the gradient of the growth rate
\begin{align}
\frac{\partial g_i}{\partial R_\alpha} = \delta_{\alpha\beta_i}\frac{\mu_{i\beta_i} k_{i\beta_i}}{(k_{i\beta_i}+R_{\beta_i})^2}\
\end{align}
to find
\begin{align}
a_i &= k_{i\beta_i}+R_{\beta_i}\\
b_{\beta_i} &= \frac{\nu_{\beta_i i}}{k_{i\beta_i}}R_{\beta_i}.
\end{align}
The competitive exclusion principle guarantees that there is at most one consumer species $i$ limited by each resource $\alpha$, which allows us to unambiguously index the functions $b_\alpha$ in this way. Aside from the strange indexing, this is the same $b_\alpha$ as in all the other resource competition models discussed so far, with effective resource weights 
\begin{align}
w_{\beta_i} = \frac{k_{i\beta_i}}{\nu_{\beta_i i}}.
\label{eq:wlie}
\end{align}
Plugging these results into eqns.~(\ref{eq:htild}) and (\ref{eq:dmod2}), we obtain an expression for $d$ that is identical to the case of substitutable resources, but with a modified supply point:
\begin{align}
\frac{\partial d}{\partial R_\alpha} &= -\frac{\tau^{-1} w_\alpha (\tilde{R}_\alpha^0 - R_\alpha)}{R_\alpha}.
\end{align}
The effective supply point is
\begin{align}
\tilde{R}_{\alpha}^0 = R_{\alpha}^0 - \tau \sum_{i,\alpha\neq \beta_i} \bar{N}_i \nu_{\alpha i} \frac{\mu_{i\beta_i} R_{\beta_i}}{k_{i\beta_i}+R_{\beta_i}}
\end{align}
with the second term accounting for the total consumption of resource $\alpha$ over a chemostat turnover time by organisms that are limited by some other resource $(\beta_i \neq \alpha)$. 

Note that the weights $w_\alpha$ in eq.~(\ref{eq:wlie}) are only defined for resources that are limiting for some species. Resources that are not limiting for any species are not subject to any constraints in the optimization, and always reach the effective supply point regardless of the values of the weights. The weights can therefore be set arbitrarily for these resources, for example by taking them all to equal 1.

\subsection*{Interactively essential resources}
Another model not discussed in the main text due to space constraints, but of interest to some readers, is the following scenario of interactively essential resources, with growth rate governed by the product of all the incoming nutrient fluxes, each following Michaelis-Menten kinetics:
\begin{align}
\frac{dN_i}{dt} &= e_i N_i \left[ \prod_{\alpha} \frac{\mu_{i\alpha} R_\alpha}{k_{i\alpha}+R_\alpha} - m_i\right]\label{eq:int1}\\
\frac{dR_\alpha}{dt} &= \tau^{-1} (R_\alpha^0 - R_\alpha) - \sum_i N_i\frac{\mu_{i\alpha} R_\alpha}{k_{i\alpha}+R_\alpha}\label{eq:int2}.
\end{align}
Comparing with the general niche theory scheme of eq.~(\ref{eq:niche1}-\ref{eq:niche2}), we identify
\begin{align}
g_i(\mathbf{R}) &= e_i \left[ \prod_{\alpha} \frac{\mu_{i\alpha} R_\alpha}{k_{i\alpha}+R_\alpha} - m_i\right]\\
q_{i\alpha}(\mathbf{R}) &= - \frac{\mu_{i\alpha} R_\alpha}{k_{i\alpha}+R_\alpha}\\
h_\alpha(\mathbf{R}) &= \tau^{-1}(R_\alpha^0-R_\alpha).
\end{align}
The gradient of the growth rate is
\begin{align}
\frac{\partial g_i}{\partial R_\alpha} &= e_i\frac{\mu_{i\alpha}k_{i\alpha}\prod_{\beta\neq\alpha} \mu_{i\beta}R_\beta}{(k_{i\alpha}+R_\alpha)^2\prod_{\gamma\neq\alpha}(k_{i\gamma}+R_\gamma)}.
\end{align}
In general, there are no functions $a_i$ and $b_\alpha$ that relate this gradient to the impact vector in the way required by Step 1 of the MEPP procedure. But if the low-density specific consumption rate $\mu_{i\alpha}/k_{i\alpha}$ is the same for all species $i$, so that we can define $w_\alpha = k_{i\alpha}/\mu_{i\alpha}$ with the left-hand side independent of $i$, we obtain:
\begin{align}
b_\alpha &= \frac{R_\alpha^2}{w_\alpha}\\
a_i &=  e_i^{-1} \prod_{\alpha} \frac{k_{i\alpha}+R_\alpha}{\mu_{i\alpha} R_\alpha}.
\end{align}
The resulting expression for $d$ is:
\begin{align}
\frac{\partial d}{\partial R_\alpha} &= -\frac{\tau^{-1} w_\alpha (R_\alpha^0 - R_\alpha)}{R_\alpha^2}.
\end{align}
Integrating this, we obtain a weighted KL divergence between the inverse resource concentrations and the inverse supply point:
\begin{align}
d(\mathbf{R}^0,\mathbf{R}) = \tau^{-1} \sum_\alpha R_\alpha^0 w_\alpha \left[ \frac{1}{R_\alpha^0} \ln \frac{1/R_\alpha^0}{1/R_\alpha} - \left(\frac{1}{R_\alpha^0}-\frac{1}{R_\alpha}\right)\right].
\end{align}

\subsection*{Type II functional response}
We begin with the dynamical equations
\begin{align}
\frac{dN_i}{dt} &= e_i N_i \left[ \sum_{\alpha} \frac{c_{i\alpha} R_\alpha}{1+\sum_{\beta} \frac{c_{i\beta}R_\beta}{J_{i\beta}}} - m_i\right]-\tau^{-1} N_i\label{eq:rothN2}\\
\frac{dR_\alpha}{dt} &= \tau^{-1} (R_\alpha^0 - R_\alpha) - \sum_i N_i \frac{c_{i\alpha} R_\alpha}{1+\sum_{\beta} \frac{c_{i\beta}R_\beta}{J_{i\beta}}}.\label{eq:rothR2}
\end{align}
Comparing with the general niche theory scheme of eq.~(\ref{eq:niche1}-\ref{eq:niche2}), we identify
\begin{align}
g_i(\mathbf{R}) &= e_i \left[ \sum_{\alpha} \frac{c_{i\alpha} R_\alpha}{1+\sum_{\beta} \frac{c_{i\beta}R_\beta}{J_{i\beta}}} - m_i\right]-\tau^{-1} \\
q_{i\alpha}(\mathbf{R}) &= - \frac{c_{i\alpha} R_\alpha}{1+\sum_{\beta} \frac{c_{i\beta}R_\beta}{J_{i\beta}}}\\
h_\alpha(\mathbf{R}) &= \tau^{-1}(R_\alpha^0-R_\alpha).
\end{align}
The gradient of the growth rate is
\begin{align}
\frac{\partial g_i}{\partial R_\alpha} &= e_i \frac{\left(1+\sum_{\beta} \frac{c_{i\beta}R_\beta}{J_{i\beta}}\right)c_{i\alpha}- \sum_{\beta}c_{i\beta}R_\beta \frac{c_{i\alpha}}{J_{i\alpha}}}{\left(1+\sum_{\beta} \frac{c_{i\beta}R_\beta}{J_{i\beta}}\right)^2}.
\end{align}
In general, there are no functions $a_i$ and $b_\alpha$ that relate this gradient to the impact vector in the way required by Step 1 of the MEPP procedure. But if the maximum uptake rates $J_{i\alpha}$ of a given consumer $i$ are the same for all resource types $\alpha$, the gradient simplifies to
\begin{align}
\frac{\partial g_i}{\partial R_\alpha} &= e_i \frac{c_{i\alpha}}{\left(1+\sum_{\beta} \frac{c_{i\beta}R_\beta}{J_{i\beta}}\right)^2}.
\end{align}
Now this can be related to $q_{i\alpha}$ in the required way, yielding
\begin{align}
a_i &= \frac{1+\sum_{\beta} \frac{c_{i\beta}R_\beta}{J_{i\beta}}}{e_i}\\
b_\alpha &= R_\alpha.
\end{align}
Since $h_\alpha$ and $b_\alpha$ are the same as for the original model with externally supplied resources and linear functional response (with $w_\alpha = 1$, because we did not need the weight parameters to fit the data of interest), the objective function is also the same. The only consequences of introducing the saturating growth law are to modify the constraint region $g_i \leq 0$ and to change the conversion factor $a_i$ required for extracting the species abundances from the Lagrange multipliers.

\section*{Appendix D: Simulation details}
All simulations and data analysis were performed in Python using the Scipy scientific computing package (\citealt{scipy}). Data and scripts (in Jupyter notebooks) to generate the figures can be downloaded from https://github.com/Emergent-Behaviors-in-Biology/mepp.

The equations parameter values for all simulations are as follows. Note that for the simulations with more than two resources, parameter values were randomly sampled. The symbol $\mathcal{U}(a,b)$ will represent a uniform probability distribution over the interval $[a,b]$, and $\mathcal{D}(\alpha)$ a Dirichlet distribution with concentration parameters all equal to $\alpha$.
\begin{itemize}
	\item {\bf Figure 2}
	\begin{itemize}
		\item (a) eq. (\ref{eq:MacN}-\ref{eq:MacR}), $c_{1\alpha} = (0.5,0.3),\, c_{2\alpha} = (0.4,0.6), \,K_\alpha = (4.8,2.85), \,r_1=r_2 = m_1 = m_2 = e_1=e_2=w_1 = w_2 =1$
		\item (b) eq. (\ref{eq:space1}-\ref{eq:space2}), $c_{1\alpha} = (0.5,0.3),\, c_{2\alpha} = (0.4,0.6), \,m_i=(0.2,0.22),\,w_i=(0.2,0.15),\,e_1=e_2=1$
		\item (c) eq. (\ref{eq:pred1}-\ref{eq:pred3}), $c_{11} = 0.5, \,c_{21} = 0.4, \,p_{11} = 0.3, \,p_{21} = 0.6, \,K_1 = 4,\,m_i=(1,0.5),\,u_1=0.5,\,r_1=w_1=e_1=e_2=1$
		\item (d) eq. (\ref{eq:chemN}-\ref{eq:chemR}), $c_{1\alpha} = (0.5,0.3),\, c_{2\alpha} = (0.4,0.6), \,R^0_\alpha= (4.8, 2.5), \,m_1=m_2=0,\,e_1=e_2=w_1 = w_2 = \tau=1$
		\item (e) eq. (\ref{eq:MacN}-\ref{eq:MacR}), $S=10,\,M=10,\,c_{i\alpha} \sim \mathcal{U}(0,1), \,K_\alpha\sim \mathcal{U}(5,6),\,r_\alpha \sim \mathcal{U}(1,2), \,m_i \sim \mathcal{U}(1,2),\,w_\alpha \sim \mathcal{U}(1,2),\,e_i=1$
		\item (f) eq. (\ref{eq:space1}-\ref{eq:space2}), $S=10,\,M=15,\,c_{i\alpha} \sim \mathcal{U}(0,1), \,m_i \sim \mathcal{U}(0.033,0.066),\,w_\alpha \sim \mathcal{U}(0.05,0.1),\,e_i=1$
	\end{itemize}
	\item {\bf Figure 3}
	\begin{itemize}
		\item (b) eq. (\ref{eq:Mi1}-\ref{eq:Mi2}), $c_{1\alpha} = (0.5,0.3),\, c_{2\alpha} = (0.4,0.6), \,D_{\alpha 1} = (0, 1),\,D_{\alpha 2} = (1,0),\,R^0_\alpha = (4.5,0.9), \,l_1=l_2 = 0.5,\,m_1 = m_2 = w_1 = w_2= e_1=e_2= \tau=1$
		\item (c) eq. (\ref{eq:Mi1}-\ref{eq:Mi2}) $S=10, \,M= 5, \,c_{i\alpha} \sim \mathcal{U}(0,1), \,D_{\alpha\beta} \sim \mathcal{D}(10),\, R^0_\alpha\sim \mathcal{U}(0,10),\,l_\alpha \sim \mathcal{U}(0,1), \,m_i \sim \mathcal{U}(1,2),\,w_\alpha \sim \mathcal{U}(1,2),\,e_i=\tau=1$
		\item (e) eq. (\ref{eq:lie1}-\ref{eq:lie2}), $\mu_{1\alpha}=(6,9),\,\mu_{2\alpha}=(8,5),\,k_{1\alpha}=k_{2\alpha}=(10,10),\,\nu_{1\alpha}=(1,0.7),\,\nu_{2\alpha} = (0.7,1),\,R^0_\alpha = (4.3,4),\,\tau=m_1=m_2=1$
		\item (f) eq. (\ref{eq:lie1}-\ref{eq:lie2}), $S=10,\,M=3,\,\mu_{i\alpha}\sim\mathcal{U}(0,30),\,k_{i\alpha}\sim\mathcal{U}(28.5,31.5),\,R_\alpha^0\sim\mathcal{U}(20,21),\,\tau=m_i=1$. The $\nu_{\alpha i}$ are generated by first sampling $\tilde{\nu}_{\alpha i}\sim (k_{i\alpha}/\mu_{i\alpha})^{-1}+\mathcal{U}(0,0.1)$, in order to increase the odds of finding stable consortia, and then normalizing with $\nu_{\alpha i} = \frac{\tilde{\nu}_{\alpha i}}{\sum_\beta \tilde{\nu}_{\beta i}}$.
	\end{itemize}
	\item {\bf Figure 4}: In panel (\emph{b}) an additional term $-\tau^{-1} N_i$ was added to eq.~(\ref{eq:II}) in the simulations, to account for the dilution of consumers. This was not required for the fitting in panel \emph{(a)}, because the growth rates were measured using the change in population size between dilutions.
	\item {\bf Figure 5}: The same additional term $-\tau^{-1} N_i$ was added to eq.~(\ref{eq:II}) for the purpose of computing the ZNGI's.
	\item {\bf Figure A1}
	\begin{itemize}
		\item (a) eq. (\ref{eq:pred1}-\ref{eq:pred3}), $S=10,\, M_R = 10,\, M_P = 10,\, c_{i\alpha}\sim \mathcal{U}(0,1),\,p_{ia}\sim \mathcal{U}(0,1),\,m_i\sim\mathcal{U}(1,2),\,u_a \sim \mathcal{U}(1,2),\,w_\alpha \sim \mathcal{U}(1,2),\,r_\alpha \sim \mathcal{U}(1,2),\,K_\alpha \sim \mathcal{U}(0,5),\,e_i=1$
		\item (b) eq. (\ref{eq:chemN}-\ref{eq:chemR}), $S=9,\, M=5,\, c_{i\alpha}\sim \mathcal{U}(0,1),\,w_\alpha \sim\mathcal{U}(1,2),\,R_\alpha^0\sim\mathcal{U}(1,7),\,\tau=e_i=1$
	\end{itemize}
\end{itemize}

\end{document}